\PassOptionsToPackage{unicode}{hyperref}
\PassOptionsToPackage{hyphens}{url}
\PassOptionsToPackage{dvipsnames,svgnames,x11names}{xcolor}
\documentclass[12pt]{article}

\usepackage{amsmath,amssymb}
\usepackage{iftex}
\ifPDFTeX
  \usepackage[T1]{fontenc}
  \usepackage[utf8]{inputenc}
  \usepackage{textcomp}
\else
  \usepackage{unicode-math}
  \defaultfontfeatures{Scale=MatchLowercase}
  \defaultfontfeatures[\rmfamily]{Ligatures=TeX,Scale=1}
\fi
\usepackage{lmodern}
\ifPDFTeX\else  
\fi
\IfFileExists{upquote.sty}{\usepackage{upquote}}{}
\IfFileExists{microtype.sty}{%
  \usepackage[]{microtype}
  \UseMicrotypeSet[protrusion]{basicmath}
}{}
\makeatletter
\@ifundefined{KOMAClassName}{%
  \IfFileExists{parskip.sty}{%
    \usepackage{parskip}
  }{%
    \setlength{\parindent}{0pt}
    \setlength{\parskip}{6pt plus 2pt minus 1pt}}
}{%
  \KOMAoptions{parskip=half}}
\makeatother
\usepackage{xcolor}
\makeatletter
\ifx\paragraph\undefined\else
  \let\oldparagraph\paragraph
  \renewcommand{\paragraph}{
    \@ifstar
      \xxxParagraphStar
      \xxxParagraphNoStar
  }
  \newcommand{\xxxParagraphStar}[1]{\oldparagraph*{#1}\mbox{}}
  \newcommand{\xxxParagraphNoStar}[1]{\oldparagraph{#1}\mbox{}}
\fi
\ifx\subparagraph\undefined\else
  \let\oldsubparagraph\subparagraph
  \renewcommand{\subparagraph}{
    \@ifstar
      \xxxSubParagraphStar
      \xxxSubParagraphNoStar
  }
  \newcommand{\xxxSubParagraphStar}[1]{\oldsubparagraph*{#1}\mbox{}}
  \newcommand{\xxxSubParagraphNoStar}[1]{\oldsubparagraph{#1}\mbox{}}
\fi
\makeatother

\usepackage{longtable,booktabs,array}
\usepackage{calc}
\usepackage{etoolbox}
\makeatletter
\patchcmd\longtable{\par}{\if@noskipsec\mbox{}\fi\par}{}{}
\makeatother
\IfFileExists{footnotehyper.sty}{\usepackage{footnotehyper}}{\usepackage{footnote}}
\makesavenoteenv{longtable}
\usepackage{graphicx}
\usepackage{tikz}
\usetikzlibrary{positioning}
\makeatletter
\def\maxwidth{\ifdim\Gin@nat@width>\linewidth\linewidth\else\Gin@nat@width\fi}
\def\maxheight{\ifdim\Gin@nat@height>\textheight\textheight\else\Gin@nat@height\fi}
\makeatother
\setkeys{Gin}{width=\maxwidth,height=\maxheight,keepaspectratio}
\makeatletter
\def\fps@figure{htbp}
\makeatother

\makeatletter
\@ifpackageloaded{caption}{}{\usepackage{caption}}
\AtBeginDocument{%
\ifdefined\contentsname
  \renewcommand*\contentsname{Table of contents}
\else
  \newcommand\contentsname{Table of contents}
\fi
\ifdefined\listfigurename
  \renewcommand*\listfigurename{List of Figures}
\else
  \newcommand\listfigurename{List of Figures}
\fi
\ifdefined\listtablename
  \renewcommand*\listtablename{List of Tables}
\else
  \newcommand\listtablename{List of Tables}
\fi
\ifdefined\figurename
  \renewcommand*\figurename{Figure}
\else
  \newcommand\figurename{Figure}
\fi
\ifdefined\tablename
  \renewcommand*\tablename{Table}
\else
  \newcommand\tablename{Table}
\fi
}
\@ifpackageloaded{float}{}{\usepackage{float}}
\floatstyle{ruled}
\@ifundefined{c@chapter}{\newfloat{codelisting}{h}{lop}}{\newfloat{codelisting}{h}{lop}[chapter]}
\floatname{codelisting}{Listing}

\makeatother
\makeatletter
\@ifpackageloaded{caption}{}{\usepackage{caption}}
\@ifpackageloaded{subcaption}{}{\usepackage{subcaption}}
\makeatother

\ifLuaTeX
  \usepackage{selnolig}
\fi
\usepackage[]{natbib}
\usepackage{bookmark}

\IfFileExists{xurl.sty}{\usepackage{xurl}}{}
\hypersetup{
  pdftitle={Neural composite likelihood estimation: simulation based inference for time series},
  pdfauthor={Grace Yan; Mark Beaumont; Dennis Prangle},
  pdfkeywords={likelihood-free inference, neural likelihood estimation, time series},
  colorlinks=true,
  linkcolor={blue},
  filecolor={Maroon},
  citecolor={Blue},
  urlcolor={Blue},
  pdfcreator={LaTeX via pandoc}}

\usepackage{todonotes} 
\usepackage{bm}
\usepackage{mathtools}
\usepackage{algorithm2e}
\usepackage{comment}
\usepackage{multirow}

\newcommand{\anon}{1}

\DeclareMathOperator{\E}{\mathbb{E}}
\DeclareMathOperator{\Var}{Var}
\DeclareMathOperator{\Hess}{Hess}

\begin{document}

\def\spacingset#1{\renewcommand{\baselinestretch}%
{#1}\small\normalsize} \spacingset{1}

\if1\anon
{
  \title{\bf Neural composite likelihood estimation: simulation based inference for time series}
  \author{Grace Yan\thanks{
    The authors gratefully acknowledge support of a PhD studentship for Grace Yan from the EPSRC Centre for Doctoral Training in Computational Statistics and Data Science (COMPASS).}\hspace{.2cm}\\
    School of Mathematics, University of Bristol\\
    and \\
    Mark Beaumont \\
    School of Biological Sciences, University of Bristol\\
    and \\
    Dennis Prangle \\
    School of Mathematics, University of Bristol}
  \maketitle
} \fi

\if0\anon
{
  \bigskip
  \bigskip
  \bigskip
  \begin{center}
    {\LARGE\bf Neural composite likelihood estimation: simulation based inference for time series}
\end{center}
  \medskip
} \fi

\bigskip

\begin{abstract}
Simulation based inference (SBI) circumvents the challenge of intractable likelihoods by using a simulator that generates data given parameter values.
For instance, neural likelihood estimation (NLE)
estimates the likelihood function by training a neural network to perform conditional density estimation on simulated data given corresponding parameters.
However such density estimation is only feasible for relatively low dimensional data.
We extend the scalability of SBI methods to a higher dimensional problem:
long sequences with a complex dependency structure.

We introduce Neural Composite Likelihood Estimation (NCLE).
This divides the sequence into smaller, equal-sized batches. Instead of training NLE to estimate the likelihood for an entire sequence, we estimate the likelihood for each batch separately.
The product of these forms an approximate composite likelihood (CL), and we perform frequentist inference using methods from the CL literature:
we get a point estimate from maximising the approximate CL and obtain confidence intervals by estimating the Godambe information matrix. 
We demonstrate the effectiveness of NCLE with experiments on time series models.
\end{abstract}

\noindent%
\vfill

\newpage
\spacingset{1.8}

\section{Introduction}

Sequence data consists of observations that are inherently dependent. They are used to capture, model, and analyse phenomena that evolve over time or in a specific order. 
Sequence data is prevalent in many applications, including finance \citep{tsay2005analysis, box2015time}, chemistry \citep{bures2023organic},
earthquake modelling \citep{stockman2024sb} and
genetics \citep{alipanahi2015predicting, beeravolu2018able}.
Recent advances have allowed the collection of longer sequences and investigation of more complex models,
but raise challenges for statistical methods.
This includes standard likelihood-based approaches,
and simulation based inference (SBI) alternatives, as we outline now.


Many classical statistical inference methods involve evaluating the likelihood $p(\mathbf{x}^{obs}|\bm{\theta})$:
the probability (or density) of the observed data $\mathbf{x}^{obs}$ given the parameters $\bm{\theta}$.
However for some complex models this is intractable: numerical likelihood evaluation is impossible or infeasibly expensive.
SBI \citep{cranmer2020frontier, deistler2025simulation} is a family of methods for \textit{likelihood-free inference} that address this problem.
It assumes access to a \textit{simulator}: a computer program that generates random samples from the model given parameters.
SBI simulates $(\bm{\theta}, \mathbf{x})$ pairs, and uses these to estimate some distribution of interest.
We focus on estimating the likelihood.
This approach dates back to \cite{diggle1984monte}, but recently has been very successful due to the development of powerful machine learning density estimation methods.
These train a neural network which can output an estimate of the likelihood $p(\mathbf{x}|\bm{\theta})$ given inputs $\mathbf{x}$ and $\bm{\theta}$.


SBI methods are still limited to relatively low dimensional data \citep{dirmeier2025simulation}.
A standard solution is to use dimension reduction from $\mathbf{x}$ to lower dimensional \emph{summary statistics} $\mathbf{s}$.
Various methods have been proposed to learn summary statistics,
either as a separate stage \citep{chen2021neural},
or via an embedding layer at the start of the neural network \citep{deistler2025simulation}.
However these methods remain expensive when $\mathbf{x}$ is sufficiently high dimensional.
Also, SBI with summary statistics still requires a large number of simulations of full data $\mathbf{x}$,
which can be infeasible when simulation of each long sequence $\mathbf{x}$ is expensive.

For these reasons, inference for long sequence data remains a challenge for SBI.
To scale up SBI to this setting, one solution is ``divide-and-conquer'' approaches.
These break up the data into smaller subsets, which we call ``batches'', and evaluate the likelihood for each batch.
We refer to these as ``sub-likelihoods''.
The sub-likelihoods are then combined by taking their product.
If data is independent across batches
then this approximates the full likelihood. 
Under dependence a \emph{composite likelihood} \citep{varin2011overview} is formed.
See Section \ref{sec:litrev} for a review of related divide-and-conquer methods.

In this paper, we present \textit{neural composite likelihood estimation} (NCLE), a composite likelihood method that leverages NLE to approximate sub-likelihoods, enabling SBI to scale to long sequence data.
The motivation is that it is much easier to estimate the likelihood of a single batch rather than the whole sequence,
reducing both the NLE error and the computational cost.
NCLE relies on assuming \emph{stationarity},
so that every simulated and observed batch has the same distribution.
We take a frequentist approach and find a NCLE point estimate by maximising the estimated composite likelihood.
For uncertainty quantification we estimate the \textit{Godambe information matrix}
to provide confidence intervals.
Our results show that NCLE, using properly selected batch lengths, improves the speed of inference compared to standard NLE
by a factor of 16 in one synthetic experiment (Section \ref{sec:AR1})
and 4 in another (Section \ref{sec:GARCH}).
In both cases the inference quality surpasses that of standard NLE.

A crucial tuning choice is the batch length.
It's important this is
sufficiently long to capture information from dependence in the sequence data at the correct scale, and
sufficiently short to avoid SBI becoming expensive and inaccurate.
We argue for using measures of autocorrelation as heuristics to find promising batch lengths,
and a simulation study to make and validate a final choice.
See Section \ref{sec:batch_length} for more discussion.


The structure of the paper is as follows. Section \ref{sec:methods} overviews existing techniques that are relevant to NCLE. Section \ref{sec:NCLE} details the NCLE algorithm, and Section \ref{sec:experiments} provides the results of using NCLE for two models:
AR(1) and GARCH(1,1).
Finally, Section \ref{sec:discussion} concludes the paper with directions for further study.

Code to reproduce our results is available at \url{https://github.com/gyanstats/ncle}.

\subsection{Related work} \label{sec:litrev}

Many divide-and-conquer methods are available for combining independent replications of data.
When the likelihood is available,
one line of work combines MCMC posterior samples from different datasets to approximate a global posterior.
This includes consensus Monte Carlo \citep{scott2016bayes},
Weierstrass refinement \citep{wang2013parallelizing},
Wasserstein barycenter methods \citep{srivastava2018scalable, ou2021scalable}
and Bayesian fusion \citep{dai2023bayesian}.
Another approach is to iteratively update approximate posteriors for each dataset, with occasional communication to help learn a global posterior.
This includes expectation propagation (EP) \citep{minka2001expectation, vehtari2020} and partitioned variational inference (PVI) \citep{bui2018partitioned}.

When the likelihood is not available,
divide-and-conquer SBI methods have previously been used 
for independent data \citep{boelts2022flexible, geffner2023compositional, linhart2024diffusion},
as well as for sequence data with a Markov model
\citep{white2015piecewise, gloeckler2025compositional}.
In the approximate Bayesian computation (ABC) literature, EP-ABC \citep{barthelme2014expectation, barthelme2018divide} replaces the intractable likelihood by a composite likelihood to obtain a variation on the ABC posterior approximation.

Composite likelihood approaches \citep{varin2011overview} can be used with more general dependence structures.
Previous work \citep{ryden1994consistent, andrieu2005online} has investigated time series using the same ``split data'' composite likelihood approximation as in the present study: equation \eqref{eq:split} later.
A major difference is that we approximate the likelihood terms using a neural surrogate (NLE).
\cite{rimella2025simulation} also combine composite likelihood and approximation using simulation, applied to hidden Markov models.
However their use of simulation is not within the framework of SBI, which we focus on.

\section{Background} \label{sec:methods}

\subsection{Notation} \label{sec:notation}

We introduce some notation used throughout the paper.
We observe a sequence 
$\mathbf{x}^{obs} = (x^{obs}_1,\ldots, x^{obs}_L)$
of length $L$ and denote a generic dataset as $\mathbf{x} = (x_1,\ldots, x_L)$,
where $x_t \in \mathbb{R}^{d_x}$.
A parameter vector is denoted as $\bm{\theta} \in \mathbb{R}^{d_\theta}$.
Although we are mainly interested in frequentist inference, we sometimes use $p(\bm{\theta})$ to denote a prior density.
Alternatively, in some contexts $p(\bm{\theta})$ represents a training density used to sample training $\bm{\theta}$ values.

We denote the computer simulator as $\texttt{simulator}(\bm{\theta}, s)$.
This generates sequence data $\mathbf{x}$ of length $s$ from our model given parameters $\bm{\theta}$.
We will consider various choices of $s$, including $s=L$ (a full length sequence) and $s=l$ (a shorter ``batch length'' for our method).
For our NCLE method later it will be important that $\texttt{simulator}$ samples from the stationary distribution,
but this is not needed for the standard SBI methods described in this section.

\subsection{Neural likelihood estimation} \label{sec:NLE}

\citet{papamakarios2019snle} propose neural likelihood estimation (NLE) as an SBI method.
It involves generating $N$ training pairs $\{ \bm{\theta}^{(i)},\mathbf{x}^{(i)} \}_{i=1}^N$.
We assume the training parameter values $\bm{\theta}^{(i)}$ are sampled from $p(\bm{\theta})$,
but they can be obtained by other methods e.g.~active learning.
The training datasets $\mathbf{x}^{(i)}$ are sampled using $\texttt{simulator}(\bm{\theta}^{(i)}, L)$.

NLE uses the training data to learn $q_{\bm{\phi}}(\mathbf{x};\bm{\theta})$,
a density estimate for $\mathbf{x}$ conditional on $\bm{\theta}$.
Here $q_{\bm{\phi}}$ is a parametric family with parameters $\bm{\phi}$ which are learned by optimisation.
We use \emph{normalising flows} \citep{kobyzev2020normalizing, papamakarios2021normalizing}, following the SBI literature.
This learns a transformation $\mathbf{x} = h_{\bm{\phi}}(\bm{z};\bm{\theta})$ where $\bm{z} \sim N(0,\mathrm{I})$
and $h$ is a neural network with weights $\bm{\phi}$, inputs $\bm{\theta}$ and an architecture allowing evaluation of $q_{\bm{\phi}}(\mathbf{x};\bm{\theta})$.
Algorithm~\ref{alg:nle} summarises the NLE method.
See \cite{deistler2025simulation} for a detailed review.


\RestyleAlgo{ruled}
\IncMargin{1em}
\begin{algorithm}[tbp]
\SetKwInOut{Input}{input}\SetKwInOut{Output}{output}
\Input{observed data $\mathbf{x}^{obs}$, training density $p(\bm{\theta})$, $\texttt{simulator}(\bm{\theta}, L)$, estimator family $q_{\bm{\phi}}(\mathbf{x};\bm{\theta})$, number of simulations $N$}
\BlankLine

\For{$i=1$ \KwTo $N$}{
sample $\bm{\theta}^{(i)} \sim p(\bm{\theta})$\\
sample $\mathbf{x}^{(i)}$ from $\texttt{simulator}(\bm{\theta}^{(i)}, L)$
}
train $\bm{\phi}$ to maximise $\sum_{i=1}^N \log q_{\bm{\phi}}(\mathbf{x}^{(i)};\bm{\theta}^{(i)})$\\
\textbf{return} approximate likelihood $q_{\bm{\phi}}(\mathbf{x}^{obs};\bm{\theta})$

\caption{Neural Likelihood Estimation (NLE)}\label{alg:nle}
\end{algorithm}\DecMargin{1em}

NLE enables various forms of statistical inference.
For approximate Bayesian inference, MCMC can be used to sample from an approximate posterior
$\hat{p}(\bm{\theta}|\mathbf{x}^{obs}) \propto p(\bm{\theta}) q_{\bm{\phi}}(\mathbf{x}^{obs};\bm{\theta})$.
For approximate frequentist inference, we can optimise $q_{\bm{\phi}}(\mathbf{x}^{obs};\bm{\theta})$ over $\bm{\theta}$, producing an approximation to the maximum likelihood estimate (MLE).

\subsubsection{Neural posterior estimation} \label{sec:NPE}

Neural posterior estimation (NPE) is a complementary SBI method to NLE \citep{papamakarios2016snpe, deistler2025simulation}.
It instead learns $q_{\bm{\phi}}(\bm{\theta} ; \mathbf{x})$: an estimate of the posterior density under prior $p(\bm{\theta})$.
This can be an easier density estimation problem, for instance when $\dim \bm{\theta} \ll \dim \mathbf{x}$.
Since our paper focuses on likelihood estimation,
we note that NPE can be used to produce an estimate of the likelihood up to proportionality: $q_{\bm{\phi}}(\bm{\theta} ; \mathbf{x}) / p(\bm{\theta})$.
In the case of a uniform prior, $q_{\bm{\phi}}(\bm{\theta} ; \mathbf{x})$ is itself a likelihood estimate (up to proportionality).
Section \ref{sec:batch_length} discusses some potential differences between using NLE and NPE to produce likelihood estimates for our method,
and Section \ref{sec:experiments} includes empirical comparisons.

\subsection{Composite likelihood} \label{sec:CL}

Composite likelihood \citep{varin2011overview}  approximates the likelihood function as the product of simpler marginal or conditional likelihoods
(which we call ``sub-likelihoods'').
%
%
We focus on the \textit{split data likelihood} \citep{ryden1994consistent}:
\begin{equation} \label{eq:split}
p_\text{CL}(\mathbf{x}^{obs};\bm{\theta}) = \prod_{b=1}^B p(\mathbf{x}_b^{obs} | \bm{\theta}).
\end{equation}
Here, $\mathbf{x}^{obs}$ is partitioned into \emph{batches}
$\mathbf{x}^{obs}_b := (x^{obs}_{(b-1)l+1}, \ldots, x^{obs}_{bl})$ for $b=1,\ldots,B$,
where $l$ is the batch length.
So $L = Bl$ and we must pick $l$ which divides $L$.

Maximising the composite likelihood gives the \emph{maximum composite likelihood estimator} (MCLE), $\hat{\bm{\theta}}_\text{CL}$.
Next we summarise typical MCLE asymptotic results.
Section \ref{sec:asymptotics} and Appendix \ref{app:regularity} discuss when these are valid.
For large $B$, $\hat{\bm{\theta}}_\text{CL}$ is often asymptotically Gaussian:
\begin{equation} \label{eq:asymptoticMCLE}
\mathbf{G}(\bm{\theta}^*)^{1/2} (\hat{\bm{\theta}}_\text{CL}-\bm{\theta}^*) \overset{d}{\rightarrow} N(\mathbf{0},\mathrm{I}),
\end{equation}
where $\bm{\theta}^*$ is the true value and $\mathbf{G}$ is the \textit{Godambe information matrix}:
\begin{align}
    \mathbf{G}(\bm{\theta}) &= \mathbf{S}(\bm{\theta})\mathbf{V}(\bm{\theta})^{-1}\mathbf{S}(\bm{\theta}), \label{eq:G_def} \\
    \text{where} \quad
    \mathbf{S}(\bm{\theta}) &= \E_{\mathbf{x}}[-\Hess \log p_\text{CL}(\mathbf{x};\bm{\theta})], \label{eq:S_def} \\
    \mathbf{V}(\bm{\theta}) &= \Var_{\mathbf{x}}[\nabla \log p_\text{CL}(\mathbf{x};\bm{\theta})]. \label{eq:V_def}
\end{align}
Here $\nabla$ and $\Hess$ are the gradient and Hessian with respect to $\bm{\theta}$.
Also, $\mathbf{S}(\bm{\theta})$ and $\mathbf{V}(\bm{\theta})$ are known as the \textit{sensitivity} and \textit{variability} matrices.
Their definitions involve the composite score function, $\nabla \log p_\text{CL}(\mathbf{x};\bm{\theta})$.
When the composite likelihood is the true likelihood then $\mathbf{S}=\mathbf{V}$,
leaving the Godambe information matrix equal to the Fisher information matrix, and the MCLE equivalent to the standard MLE.

\subsubsection{Confidence interval calculations} \label{sec:ci}

To estimate $\mathbf{S}$ and $\mathbf{V}$, the expectations in equations \eqref{eq:S_def} and \eqref{eq:V_def} could be approximated using Monte Carlo.
However this requires simulating many long $\mathbf{x}$ sequences.
Evaluation of Hessian matrices can also sometimes present difficulties, although for our later applications this would be relatively cheap by automatic differentiation.

Therefore, we calculate confidence intervals using a result of \citet{rimella2025simulation}
see their Appendix B.3.
The derivation is based on the first and second Bartlett identities:
later we comment on their validity for our method.
The result is that \eqref{eq:S_def} and \eqref{eq:V_def} can be approximated in terms of composite scores for each batch,
\begin{align}
    \mathbf{S}(\bm{\theta}) &\approx \sum_{b=1}^B \E_{\mathbf{x}}[\nabla \log p(\mathbf{x}_b|\bm{\theta}) \nabla (\log p(\mathbf{x}_b|\bm{\theta}))^\top],
    \label{eq:S_approx1} \\
    \mathbf{V}(\bm{\theta}) &\approx \sum_{b=1}^B \sum_{b'=1}^B \E_{\mathbf{x}}[\nabla \log p(\mathbf{x}_b|\bm{\theta}) \nabla (\log p(\mathbf{x}_{b'}|\bm{\theta}))^\top].
    \label{eq:V_approx1}
\end{align}
Taking Monte Carlo estimates of the expectations and evaluating at the MCLE gives
\begin{align}\label{eq:S_approx}
        \mathbf{S}(\hat{\bm{\theta}}_\text{CL}) &\approx \frac{1}{n} \sum_{i=1}^n \sum_{b=1}^B \nabla \log p(\mathbf{x}^{(i)}_b|\hat{\bm{\theta}}_\text{CL}) \nabla (\log p(\mathbf{x}^{(i)}_b|\hat{\bm{\theta}}_\text{CL}))^\top, \\ \label{eq:V_approx}
        \mathbf{V}(\hat{\bm{\theta}}_\text{CL}) &\approx \frac{1}{n} \sum_{i=1}^n \sum_{b=1}^B \sum_{b'=1}^B \nabla \log p(\mathbf{x}^{(i)}_b|\hat{\bm{\theta}}_\text{CL}) \nabla (\log p(\mathbf{x}^{(i)}_{b'}|\hat{\bm{\theta}}_\text{CL}))^\top,
\end{align}
where $(\mathbf{x}^{(i)}_1, \mathbf{x}^{(i)}_2, \ldots, \mathbf{x}^{(i)}_B)$
is an output of $\texttt{simulator}(\hat{\bm{\theta}}_\text{CL}, L)$
and the simulations for $i=1,2,\ldots,n$ are independent.

Using \eqref{eq:G_def}, \eqref{eq:S_approx} and \eqref{eq:V_approx} gives $\widehat{\mathbf{G}}(\hat{\bm{\theta}}_\text{CL})$, approximating $\mathbf{G}(\bm{\theta}^*)$.
Let $\mathbf{g}$ be the vector of diagonal elements of $\widehat{\mathbf{G}}(\hat{\bm{\theta}}_\text{CL})^{-1}$.
Then Wald-type approximate $100(1-\alpha)\%$ confidence interval bounds for the $j$th parameter are
$\hat{\theta}_{\text{CL},j} \pm z_{\alpha/2} g_j^{1/2}$,
where $z_{\alpha/2}$ is the standard normal quantile.

\subsubsection{Asymptotic theory} \label{sec:asymptotics}

Standard composite likelihood asymptotics
(see \citealp{varin2011overview} for a review)
considers independent replicated data.
In our notation these are datasets
$\mathbf{x}^{obs,j} = (x_1^{obs,j},\ldots, x_L^{obs,j})$
for $j=1,2,\ldots,m$.
There is a composite likelihood $p_\text{CL}(\mathbf{x}^{obs,j};\bm{\theta})$
for each $j$.
The overall composite likelihood is
$\prod_{j=1}^m p_\text{CL}(\mathbf{x}^{obs,j};\bm{\theta})$, and
the asymptotics hold for large $m$ (number of replications)
and fixed $L$ (dataset length).

We consider instead the setting of large $L$ and $m=1$: a single long sequence.
Various authors prove asymptotic results in this setting for particular choices of model and composite likelihood.
For the split data likelihood setting, \citet{ryden1994consistent}
proves consistency and asymptotic normality as in \eqref{eq:asymptoticMCLE}
for the case of hidden Markov models
(i.e.~with discrete hidden states).
\citet{andrieu2005online} prove similar results
for state space models (i.e.~with continuous hidden states).

It's common that the entries of $\Var(\hat{\bm{\theta}}_\text{CL})$ are $O(1/L)$.
However other rates of convergence are possible.
For instance in population genetics it's conjectured that the variance is sometimes of order $\frac{1}{L}\log L$ due to long range dependence \citep{wiuf2006consistency, larribe2011composite}.

Appendix \ref{app:regularity} further discusses the above results and the regularity conditions needed.
It additionally describes how the time series models we consider as examples later do not exactly meet the regularity conditions.
Therefore it's important to empirically validate the asymptotic results.

\section{Neural composite likelihood estimation}
\label{sec:NCLE}

This section introduces our method, \textit{neural composite likelihood estimation} (NCLE), a combination of NLE and composite likelihood.

As in Section \ref{sec:notation},
we observe a sequence 
$\mathbf{x}^{obs} = (x^{obs}_1,\ldots, x^{obs}_L)$
of length $L$
and wish to infer parameters $\bm{\theta}$ under a particular model.
We assume access to $\texttt{simulator}(\bm{\theta},s)$,
which can sample a sequence of any length $s$ under $\bm{\theta}$.
We now also assume this samples from the model's stationary distribution.

\begin{figure}[tbp]
    \centering
    \includegraphics[width=0.7\linewidth]{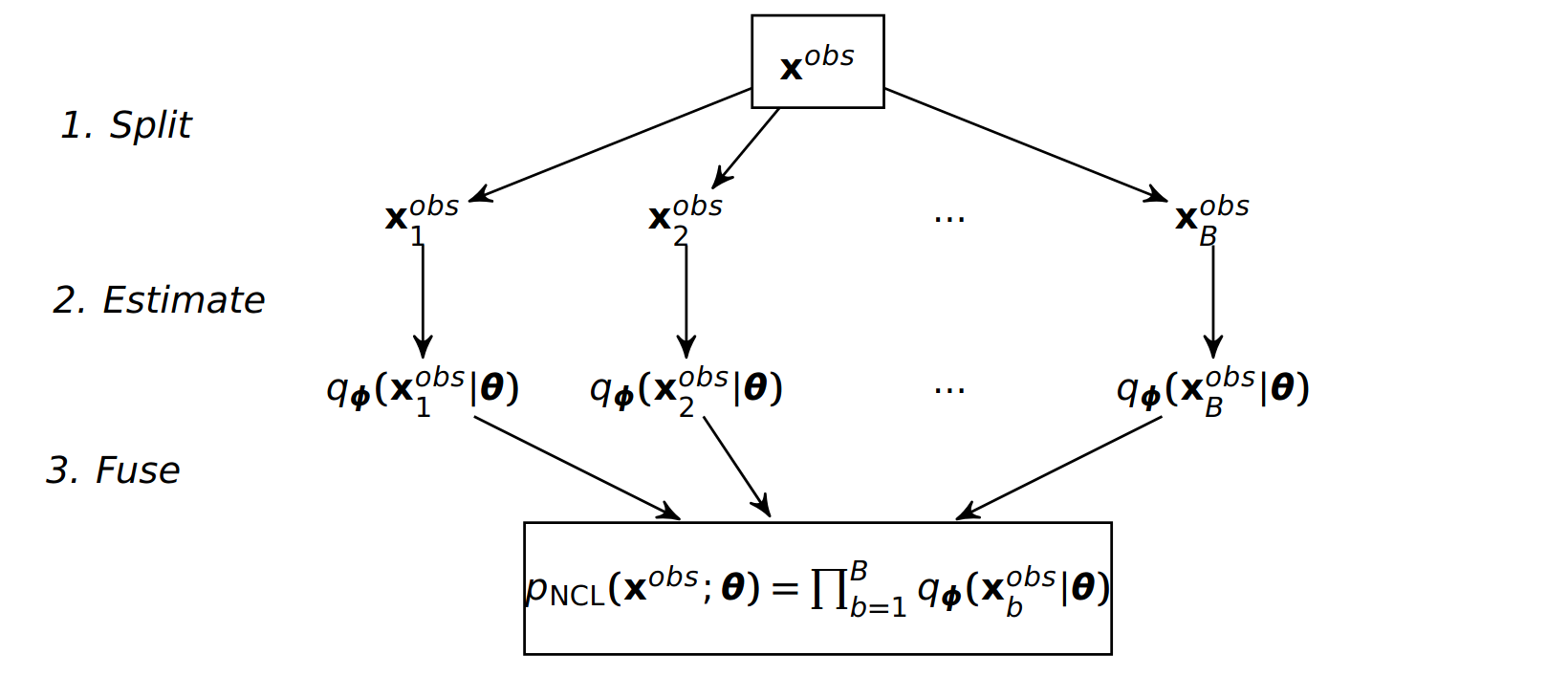}
    \caption{Neural composite likelihood steps.
    (1) Split the observed sequence (of length $L$) into $B$ batches (each of length $l$).
    (2) Approximate batch sub-likelihoods using neural density estimation.
    (3) Multiply the sub-likelihoods.}
    \label{fig:ncle_diagram}
\end{figure}

In NCLE we fix a batch length $l$, and sample $N$ training pairs 
$\{ \bm{\theta}^{(i)}, \mathbf{x}_1^{(i)} \}_{i=1}^N$,
where $\mathbf{x}_1^{(i)}$\footnote{
    The subscript $1$ indicates this is a single batch of data,
    and we're not generating $B$ batches for each $\bm{\theta}^{(i)}$.
}
is generated from $\texttt{simulator}(\bm{\theta}^{(i)},l)$.
These are used to train $q_{\bm{\phi}}(\mathbf{x}_1;\bm{\theta})$ by SBI
(either NLE or NPE -- see Sections \ref{sec:NLE} and \ref{sec:NPE}.)
We use this to approximate all the sub-likelihoods: $q_{\bm{\phi}}(\mathbf{x}^{obs}_b ; \bm{\theta}) \approx p(\mathbf{x}^{obs}_b| \bm{\theta})$. Their product gives the \emph{neural composite likelihood} (NCL):
\begin{align}\label{eq:ncl_eq1}
    p_\text{NCL}(\mathbf{x}^{obs}; \bm{\theta}) = \prod_{b=1}^B q_{\bm{\phi}}(\mathbf{x}^{obs}_b;\bm{\theta}).
\end{align}
Figure \ref{fig:ncle_diagram} gives a conceptual sketch of NCL.

An important requirement for this approach is stationarity.
We assume (1) the model is stationary
(2) observations are generated from the stationary distribution
(3) \texttt{simulator} simulates from the stationary distribution.
This is essential so that every $\mathbf{x}^{obs}_b$ and every simulated $\mathbf{x}^{(i)}_1$
has the same distribution.

Learning the likelihood of a data batch $\mathbf{x}^{obs}_b$ can be challenging when its length $l$ or dimensionality $d_x$ is large.
So a variation is to use low dimensional summary statistics
$\mathbf{s}^{obs}_b := S(\mathbf{x}^{obs}_b)$
for some function $S$.
Then SBI is trained on simulated $\mathbf{s}^{(i)} = S(\mathbf{x}^{(i)}_{1})$ values.
The resulting neural composite likelihood is
\begin{align}\label{eq:ncl_eq2}
    p_\text{NCL}(\mathbf{s}^{obs}; \bm{\theta}) = \prod_{b=1}^B q_{\bm{\phi}}(\mathbf{s}^{obs}_b;\bm{\theta}).
\end{align}

Algorithm \ref{alg:ncle} summarises our NCLE algorithm.
This starts by simulating training data and SBI training.
The resulting NCL is maximised to give $\hat{\bm{\theta}}_\text{NCL}$, an estimate of the MCLE.
Then confidence intervals are produced as in Section \ref{sec:ci} using an estimate of the Godambe information matrix.
Algorithm \ref{alg:gim} details the steps to estimate the latter.
This involves calculating gradients of $\log q_{\bm{\phi}}(\mathbf{s}^{obs}_b;\bm{\theta})$, which can be done by automatic differentiation
(since $q_{\bm{\phi}}$ is a normalising flow).
Both algorithms are described in terms of summary statistics.
To use the raw data, simply set $S(\mathbf{x})=\mathbf{x}$.




\IncMargin{1em}
\begin{algorithm}[tbp]
\SetKwInOut{Input}{input}\SetKwInOut{Output}{output}
\Input{observed data $\mathbf{x}^{obs}$, choice of summary statistics, training density $p(\bm{\theta})$, batch length $l$, $\texttt{simulator}(\bm{\theta}, s)$, training set size $N$, significance level $\alpha$}
\BlankLine
\For{$i=1$ \KwTo $N$}{
sample $\bm{\theta}^{(i)} \sim p(\bm{\theta})$ \\
simulate $\mathbf{x}^{(i)}_1 \sim \texttt{simulator}(\bm{\theta}^{(i)}, l)$ \\
calculate $\mathbf{s}^{(i)} = S(\mathbf{x}^{(i)}_1)$
}
train $\bm{\phi}$ to maximise $\sum_{i=1}^N \log q_{\bm{\phi}}(\mathbf{s}^{(i)};\bm{\theta}^{(i)})$\\
find estimated MCLE $\hat{\bm{\theta}}_\text{NCL}$ by maximising $p_\text{NCL}(\mathbf{s}^{obs};\bm{\theta}) = \prod_{b=1}^B q_{\bm{\phi}}(\mathbf{s}^{obs}_b;\bm{\theta})$\\
estimate the Godambe information matrix $\widehat{\mathbf{G}}(\hat{\bm{\theta}}_\text{NCL})$ using Algorithm \ref{alg:gim} \\
calculate $\mathbf{g}$, the vector of diagonal elements of $\widehat{\mathbf{G}}(\hat{\bm{\theta}}_\text{NCL})^{-1}$
\\
\textbf{return} estimated MCLE, $\hat{\bm{\theta}}_\text{NCL}$, and confidence interval bounds
$\hat{\theta}_{\text{NCL},j}\;\pm\; z_{\alpha/2}\, g_j^{1/2}$
for $j=1,2,\ldots,d_\theta$.

\caption{Neural Composite Likelihood Estimation (NCLE)}\label{alg:ncle}
\end{algorithm}\DecMargin{1em}

\IncMargin{1em}
\begin{algorithm}[tbp]
\SetKwInOut{Input}{input}\SetKwInOut{Output}{output}
\Input{MCLE $\hat{\bm{\theta}}_\text{NCL}$, trained density estimator $q_{\bm{\phi}}(\mathbf{s};\bm{\theta})$, $\texttt{simulator}(\bm{\theta}, s)$, number of batches $B$, batch length $l$, number of Monte Carlo simulations $n$}
\BlankLine
\For{$i=1$ \KwTo $n$}{
simulate $\mathbf{x}^{(i)} \sim \texttt{simulator}(\hat{\bm{\theta}}_\text{NCL}, L)$ \\
split $\mathbf{x}^{(i)}$ into $B$ batches of length $l$: $(\mathbf{x}^{(i)}_1, \mathbf{x}^{(i)}_2, \ldots, \mathbf{x}^{(i)}_B)$ \\
\BlankLine
\For{$b=1$ \KwTo $B$}{
calculate summary statistics $\mathbf{s}^{(i)}_b = S(\mathbf{x}^{(i)}_b)$ \\
compute score $u^{(i)}_b(\hat{\bm{\theta}}_\text{NCL}) = \nabla_{\bm{\theta}} \log q_{\bm{\phi}}(\mathbf{s}^{(i)}_b; \hat{\bm{\theta}}_\text{NCL})$ 
}
}
\BlankLine
estimate sensitivity matrix: 
$\widehat{\mathbf{S}}(\hat{\bm{\theta}}_\text{NCL}) = \frac{1}{n} \sum_{i=1}^n \sum_{b=1}^B u^{(i)}_b(\hat{\bm{\theta}}_\text{NCL}) \, u^{(i)}_b(\hat{\bm{\theta}}_\text{NCL})^\top$ \\
\BlankLine
compute sum over batches: $u^{(i)}(\hat{\bm{\theta}}_\text{NCL}) = \sum_{b=1}^B u^{(i)}_b(\hat{\bm{\theta}}_\text{NCL})$ for $i=1,\ldots,n$ \\
\BlankLine
estimate variability matrix: 
$\widehat{\mathbf{V}}(\hat{\bm{\theta}}_\text{NCL}) = \frac{1}{n} \sum_{i=1}^n u^{(i)}(\hat{\bm{\theta}}_\text{NCL}) \, u^{(i)}(\hat{\bm{\theta}}_\text{NCL})^\top$ \\
\BlankLine
compute Godambe matrix: $\widehat{\mathbf{G}}(\hat{\bm{\theta}}_\text{NCL}) = \widehat{\mathbf{S}}(\hat{\bm{\theta}}_\text{NCL}) \, \widehat{\mathbf{V}}(\hat{\bm{\theta}}_\text{NCL})^{-1} \, \widehat{\mathbf{S}}(\hat{\bm{\theta}}_\text{NCL})$ \\
\BlankLine
\textbf{return} estimated Godambe information matrix $\widehat{\mathbf{G}}(\hat{\bm{\theta}}_\text{NCL})$
\caption{Godambe Information Matrix Estimation}\label{alg:gim}
\end{algorithm}\DecMargin{1em}

Finally, recall that the Godambe estimation approach in Section \ref{sec:ci} is based on Bartlett identities.
These rely on $q_{\bm{\phi}}(\mathbf{s};\bm{\theta})$ being a sufficiently good approximation to $p(\mathbf{s}|\bm{\theta})$.
Hence if there is sufficient SBI error, then the Godambe estimates may be inaccurate and produce poor confidence intervals.

\subsection{Batch length} \label{sec:batch_length}

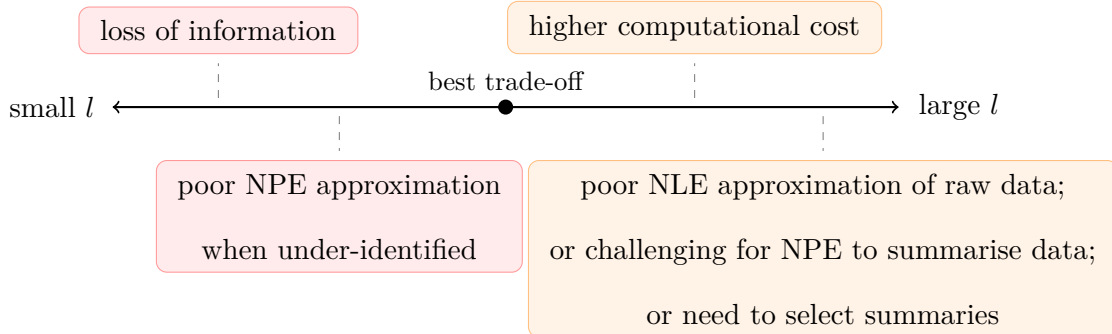
\begin{figure}[htbp]
\centering
\begin{tikzpicture}[
  box/.style={
    draw, rounded corners=4pt, font=\small,
    inner xsep=8pt, inner ysep=5pt
  },
  smallbox/.style={box, fill=red!8, draw=red!40},
  largebox/.style={box, fill=orange!10, draw=orange!50},
  every node/.style={align=center}
]

\draw[<->, thick] (-5.2,0) -- (5.2,0);
\node[font=\small] at (-6, 0) {small $l$};
\node[font=\small] at ( 6, 0) {large $l$};

\node[circle, fill=black, inner sep=2pt] at (0,0) {};
\node[font=\footnotesize] at (0, 0.35) {best trade-off};

\node[smallbox, above=0.7cm of {(-3.8,0)}] (loss)
  {loss of information};
\draw[dashed, gray] (-3.8, 0.12) -- (-3.8, 0.58);

\node[smallbox, below=0.7cm of {(-2.2,0)}] (npe)
  {poor NPE approximation\\when under-identified};
\draw[dashed, gray] (-2.2, -0.12) -- (-2.2, -0.58);

\node[largebox, above=0.7cm of {(2.5,0)}] (nle)
  {higher computational cost};
\draw[dashed, gray] (2.5, 0.12) -- (2.5, 0.58);

\node[largebox, below=0.7cm of {(4.2,0)}] (cost)
  {poor NLE approximation of raw data; \\ or challenging for NPE to summarise data; \\ or need to select summaries};
\draw[dashed, gray] (4.2, -0.12) -- (4.2, -0.58);

\end{tikzpicture}
\caption{Conceptual summary of batch length trade-offs.}
\label{fig:batch_length}
\end{figure}

Batch length $l$ is an important tuning choice for our method.
Below we list several effects that $l$ can have, then discuss how to tune it.
Figure \ref{fig:batch_length} summarises the trade-offs.

Small $l$ can cause \emph{loss of information} about long-range dependencies available in the full likelihood.
It can also cause poor \emph{NPE approximation quality}:
if $\bm{\theta}$ is poorly identified given $\mathbf{x}_b$
then $p(\bm{\theta}|\mathbf{x}_b)$ can be complicated and hard to approximate.
This can make likelihood estimation using NPE difficult.

Large $l$ can cause poor \emph{NLE approximation quality}.
Since $\mathbf{x}_b$ is high dimensional,
$p(\mathbf{x}_b|\bm{\theta})$ is a challenging density estimation problem.
Using low dimensional summary statistics $S(\mathbf{x}_b)$ avoids this problem,
but there is instead the well known problem of \emph{choosing informative summaries} \citep{chen2021neural}.
NPE instead performs density estimation of $\bm{\theta}$, so is potentially more scalable to the dimension of $\mathbf{x}_b$ \citep{deistler2025simulation}.
However for large $l$, estimating $p(\bm{\theta}|\mathbf{x}_b)$ still requires processing the high dimensional input $\mathbf{x}_b$.
This is a challenging \emph{data summarisation} task which may require a large amount of training data.
Large $l$ also gives high \emph{SBI training cost} as simulating $(\bm{\theta}, \mathbf{x}_b)$ pairs becomes expensive.


Prior work with analytical CL suggests picking $l$ large enough that the inference has stabilised,
with \cite{andrieu2005online} mentioning stability of the MCLE,
and \cite{ryden1994consistent} stability of confidence interval widths.
For NCLE this approach is more difficult as performance can decay for large $l$ due to SBI error. 

Given all the considerations above, we recommend testing NCLE on simulated data to investigate and validate the choice of $l$.
See the end of Sections \ref{sec:AR1comments}, \ref{sec:garch_comments} for example.

As an initial heuristic, we also look for a value of $l$ giving low dependence between batches,
meaning there is hopefully little loss of information.
We use simulated data to explore generic dependence diagnostics such as autocorrelation (in Sections \ref{sec:ar1_batch_length}, \ref{sec:GARCH_batch_length}).

\section{Experiments} \label{sec:experiments}

This section illustrates NCLE with experiments.
First Section \ref{sec:diagnostics} describes some diagnostics we use to report our results.
Then we conduct experiments on time series models: AR(1) (Section \ref{sec:AR1}) and GARCH(1,1) (Section \ref{sec:GARCH}).
These both use synthetic data.

We investigate, amongst other things, different choices of batch length $l$ in NCLE.
Taking $l=L$ gives a single batch, so here NCLE reduces to standard SBI.
So our results typically include a comparison to standard SBI,
which can be found by looking at the $l=L$ case.

Throughout we used the \texttt{sbi-0.22.0} toolbox \citep{tejero-cantero2020sbi} to train NLE and NPE using masked autoregressive flows \citep{papamakarios2017masked}.
To obtain the MCLE in Algorithm \ref{alg:ncle},
we used Adam with learning rate $0.001$ for the AR(1) experiments and L-BFGS-B with line search for the GARCH(1,1) experiments.

\subsection{Diagnostics} \label{sec:diagnostics}

This section introduces several diagnostics we use to report our results.

To compare different likelihood estimates (before any use of Godambe information) we often plot \emph{normalised likelihoods}. For our trained neural composite likelihood (NCL), this is
\begin{equation}\label{eq:normalised_ncl}
\tilde{p}_\text{NCL}(\mathbf{x}^{obs};\bm{\theta}) = p_\text{NCL}(\mathbf{x}^{obs};\bm{\theta}) / \max_{\bm{\vartheta}} p_\text{NCL}(\mathbf{x}^{obs};\bm{\vartheta}).
\end{equation}
When we have access to the exact analytic composite likelihood, we define the normalised composite likelihood $\tilde{p}_\text{CL}(\mathbf{x}^{obs};\bm{\theta})$ similarly, and include it in plots.
Normalised likelihood plots allow comparison of likelihood shapes even when their magnitudes are on different scales.
For scalar parameters we plot normalised likelihoods for evenly-spaced parameter values within the main support of $p(\bm{\theta})$.
Otherwise we plot the normalised likelihood for each parameter marginally, holding the remaining parameters fixed at their true values.

To measure the difference between the analytical composite log-likelihood (when available) and a neural composite log-likelihood, we calculate root mean squared error (RMSE)
\begin{equation}\label{eq:rmse}\sqrt{\frac{1}{k}\sum_{i=1}^k \Big[ \log p_\text{CL}(\mathbf{x}^{obs}|\bm{\theta}^{(i)})-\log p_\text{NCL}(\mathbf{x}^{obs}|\bm{\theta}^{(i)}) \Big]^2},\end{equation}
using $k$ evenly spaced $\bm{\theta}^{(i)}$ values from the main support of $p(\bm{\theta})$.
We calculate the RMSE for each parameter marginally, holding the remaining parameters fixed at their true values. 

We calculate the Godambe matrix using Algorithm \ref{alg:gim}.
In the single batch case ($l=L$ and $B=1$),
this estimates the Fisher information matrix.

In our experiments,
we generate a single synthetic observed dataset $\mathbf{x}^{obs}$ under known parameter values.
This is reused in all relevant plots and diagnostics
i.e.~all those based on a single dataset.
Also we use the same $p_\text{NCL}$ (trained on the same training samples) across plots where possible
i.e.~when they have the same tuning choices.

To assess the quality of uncertainty quantification, we simulate 200 datasets under the same true parameter values, and use Algorithm \ref{alg:ncle} to get MCLEs and 95\% CIs.
This allows us to estimate 95\% coverage for each parameter, as well as MCLE mean and variance.
For any particular value of $l$, we reuse the same 200 datasets while varying other tuning choices.
However, we generate new datasets when $l$ is changed. 

\subsection{AR(1) model} \label{sec:AR1}

The AR(1) model takes
$x_t = \theta x_{t-1} + \epsilon_t$ for $t \geq 2$
with $\epsilon_t \sim N(0,1)$ (all independent).
The autoregressive coefficient $\theta$ is the parameter of interest.
We assume $\theta \in (-1,1)$ to ensure stationarity,
and for $x_1$ we use the stationary distribution
$N(0,\frac{1}{1-\theta^2})$.
The likelihood is:
\begin{equation}
    p(\mathbf{x}|\theta) = \Phi\left(x_1; 0,\frac{1}{1-\theta^2} \right) \prod_{t=2}^{L} \Phi(x_t; \theta x_{t-1}, 1),
\end{equation}
where $\Phi(x;\mu,\sigma^2)$ is a $N(\mu,\sigma^2)$ density.
See \citet{hamilton1994time} for more background.
Appendix \ref{app:regularity} discusses the composite likelihood regularity conditions,
noting that existing theory doesn't cover this simple model, so empirical validation is needed.

Our synthetic observations were simulated under $\theta^*=0.8$ and we used $p(\theta)$ matching the uniform distribution $U(-1,1)$.
We took sequence length $L=1000$ and
batch length $l \in \{2,5,10,20,100,500,1000\}$.
Taking $l=1000$ means the NCLE is equivalent to standard NLE, and the analytical composite likelihood is the true likelihood.

We investigate NCLE using the raw data -- i.e.~$S(\mathbf{x}_1)=\mathbf{x}_1$ --
and also using sufficient summary statistics,
\begin{equation}
\label{eq:ar1_sufficientSS}
    \mathcal{S}_2(\mathbf{x}_1) = \left(\sum_{t=2}^{l-1} x_t^2, \quad \sum_{t=2}^l x_{t-1}x_t \right).
\end{equation}
For the derivation of these sufficient statistics, see Appendix \ref{sec:ar1_deriv}.
We used $N=10^4$ and $N=10^5$ training samples for the raw data, and $N=10^5$ for $\mathcal{S}_2$.

\subsubsection{Batch length heuristic} \label{sec:ar1_batch_length}

As discussed in Section \ref{sec:batch_length}, some guidance on a suitable batch length can be found using the autocorrelation structure.
For an AR(1) process, the autocorrelation function (ACF) at lag \(k\) is \(C(k)=\theta^k\) \citep{hamilton1994time}.
Solving $C(k)=c$ with $c=0.1$ gives \(k=10.3\) under $\theta^*$, suggesting that $l=10$ is a reasonable choice to capture short-range dependence in the data.
For our synthetic data $\mathbf{x}^{obs}$,
the empirical ACF drops below $c$ at a larger value, $k=23$.
Our results investigate how well these values behave as choices of batch length in practice.

\subsubsection{Results}

\begin{figure}[tbp]
    \centering
    \begin{subfigure}{0.325\textwidth}
        \centering
        \includegraphics[width=1\textwidth]{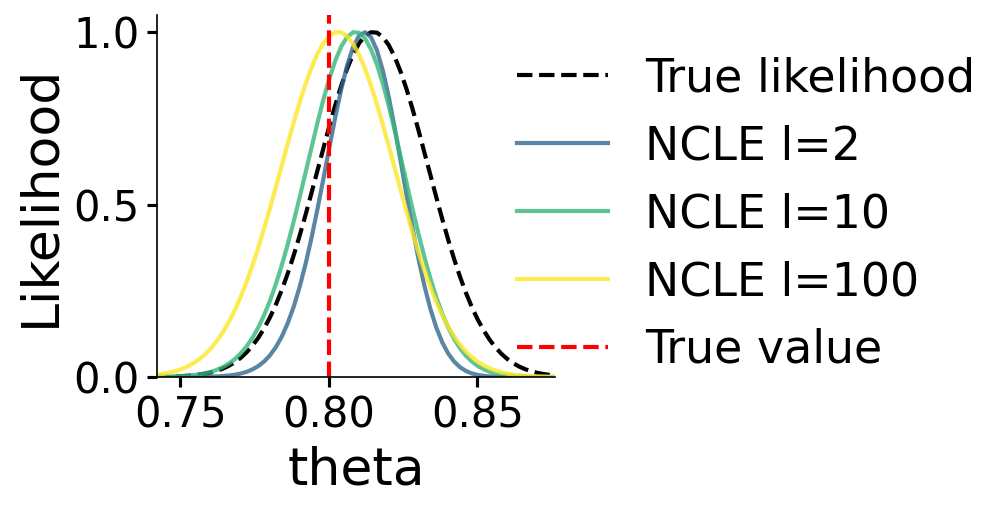}
        \phantomcaption  
        \put(-160,80){\textbf{(a)}} 
        \label{fig:ar1_ncls}
    \end{subfigure}
    \begin{subfigure}{0.325\textwidth}
        \centering
        \includegraphics[width=1\textwidth]{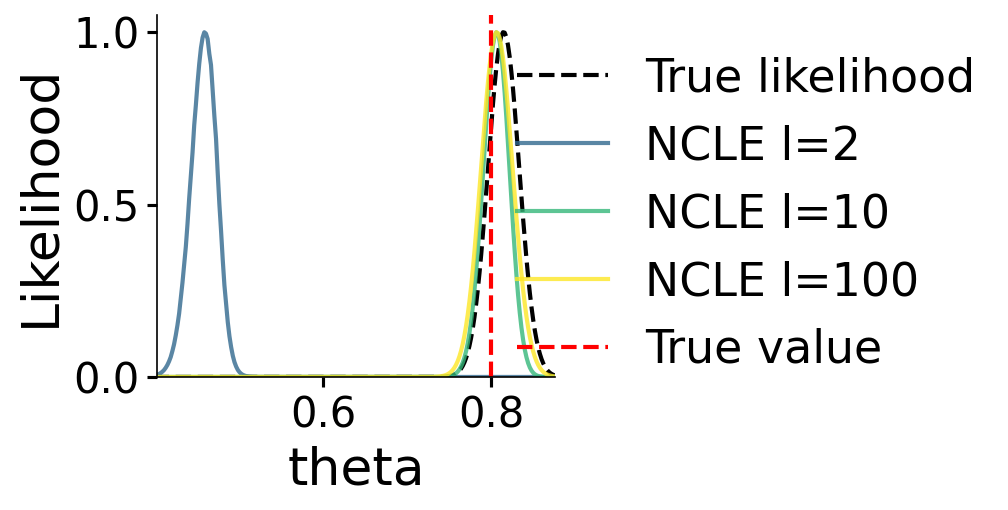}
        \phantomcaption  
        \put(-160,80){\textbf{(b)}}
        \label{fig:sufficientSS_ncls}
    \end{subfigure}
    \hfill
    \begin{subfigure}{0.325\textwidth}
        \centering
        \includegraphics[width=1\linewidth]{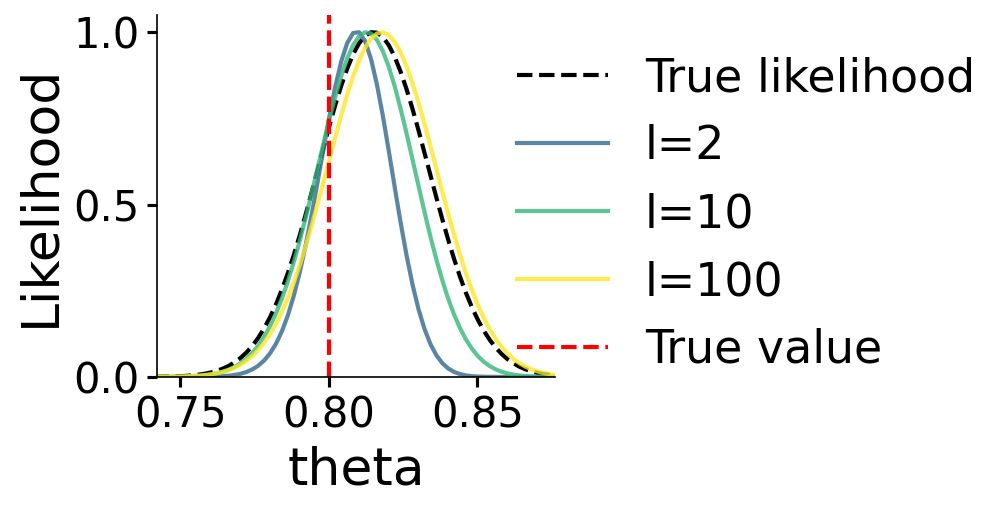}
        \put(-150,80){\textbf{(c)}}
        \phantomcaption  
        \label{fig:ar1_cls}
    \end{subfigure}
    \begin{subfigure}[t]{0.31\textwidth}
        \centering
        \begin{subfigure}{\textwidth}
            \centering
            \includegraphics[width=1\linewidth]{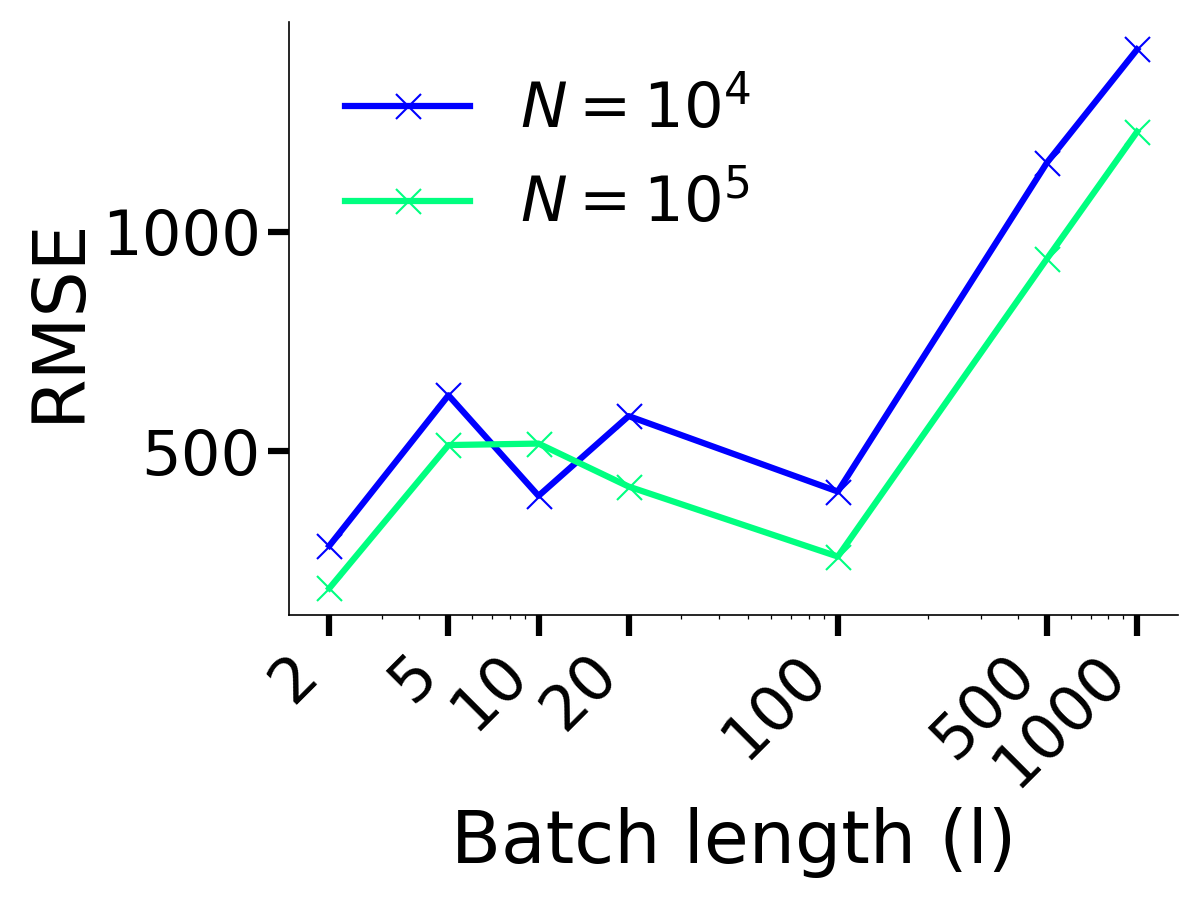}
            \put(-130,100){\textbf{(d)}}
            \phantomcaption
            \label{fig:ar1_rmses}
        \end{subfigure}
        \begin{subfigure}{\textwidth}
            \centering
            \includegraphics[width=1\linewidth]{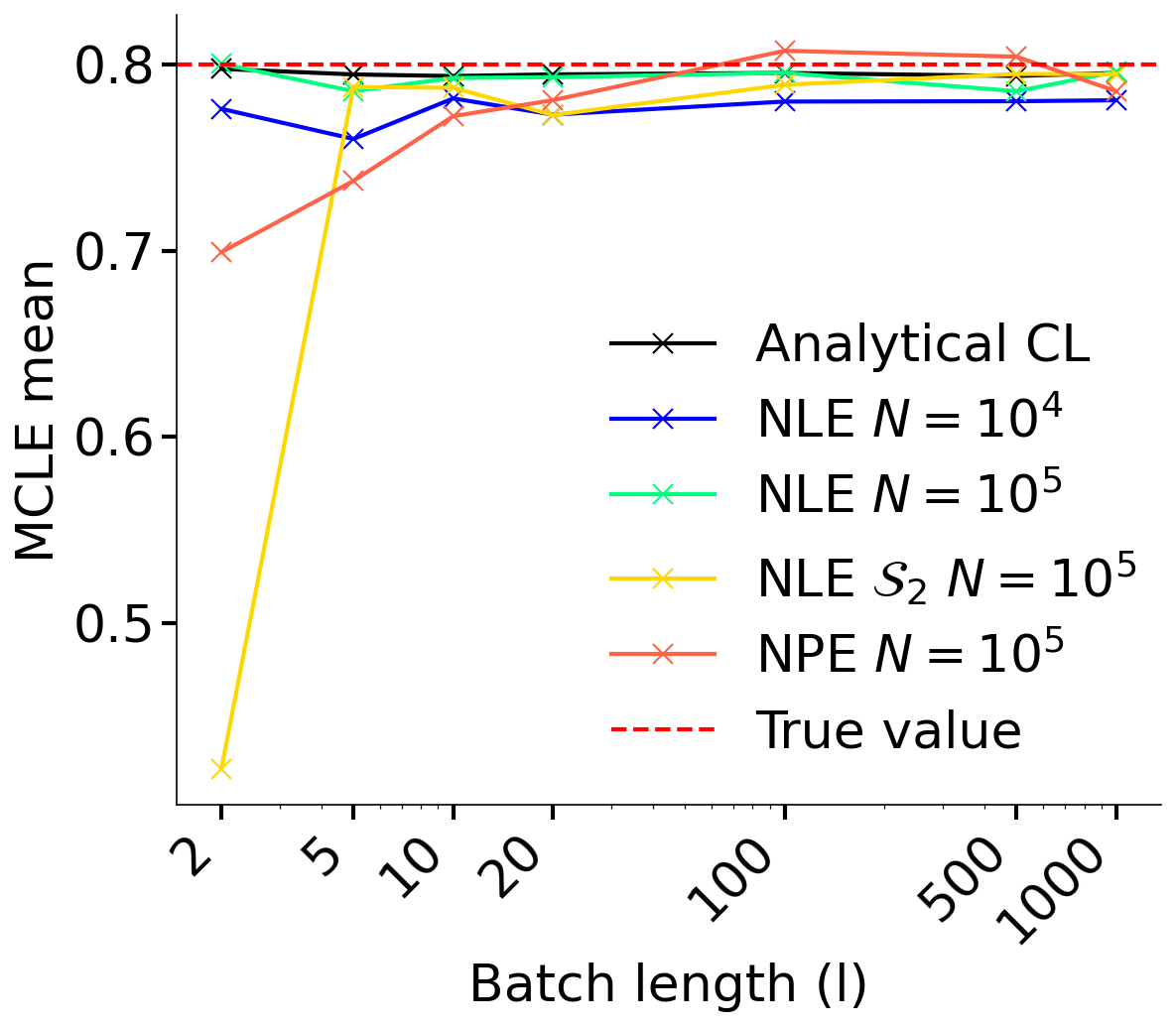}
            \put(-140,125){\textbf{(g)}}
            \phantomcaption
            \label{fig:ar1_ci_widths}
        \end{subfigure}
    \end{subfigure}
    \hfill
    \begin{subfigure}[t]{0.31\textwidth}
        \centering
        \begin{subfigure}{\textwidth}
            \centering
            \includegraphics[width=1\linewidth]{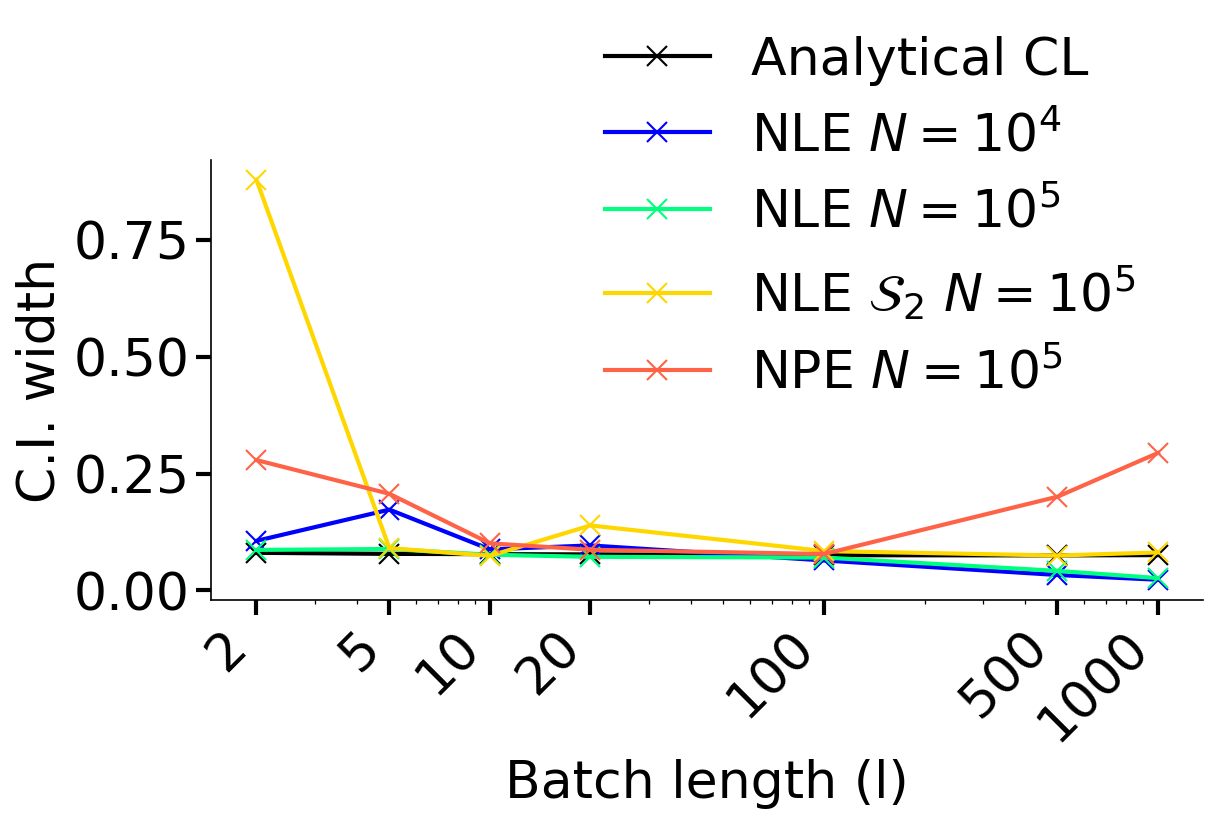}
            \put(-140,90){\textbf{(e)}}
            \phantomcaption  
            \label{fig:ar1_times}
        \end{subfigure}
        \begin{subfigure}{\textwidth}
            \centering
            \includegraphics[width=1\linewidth]{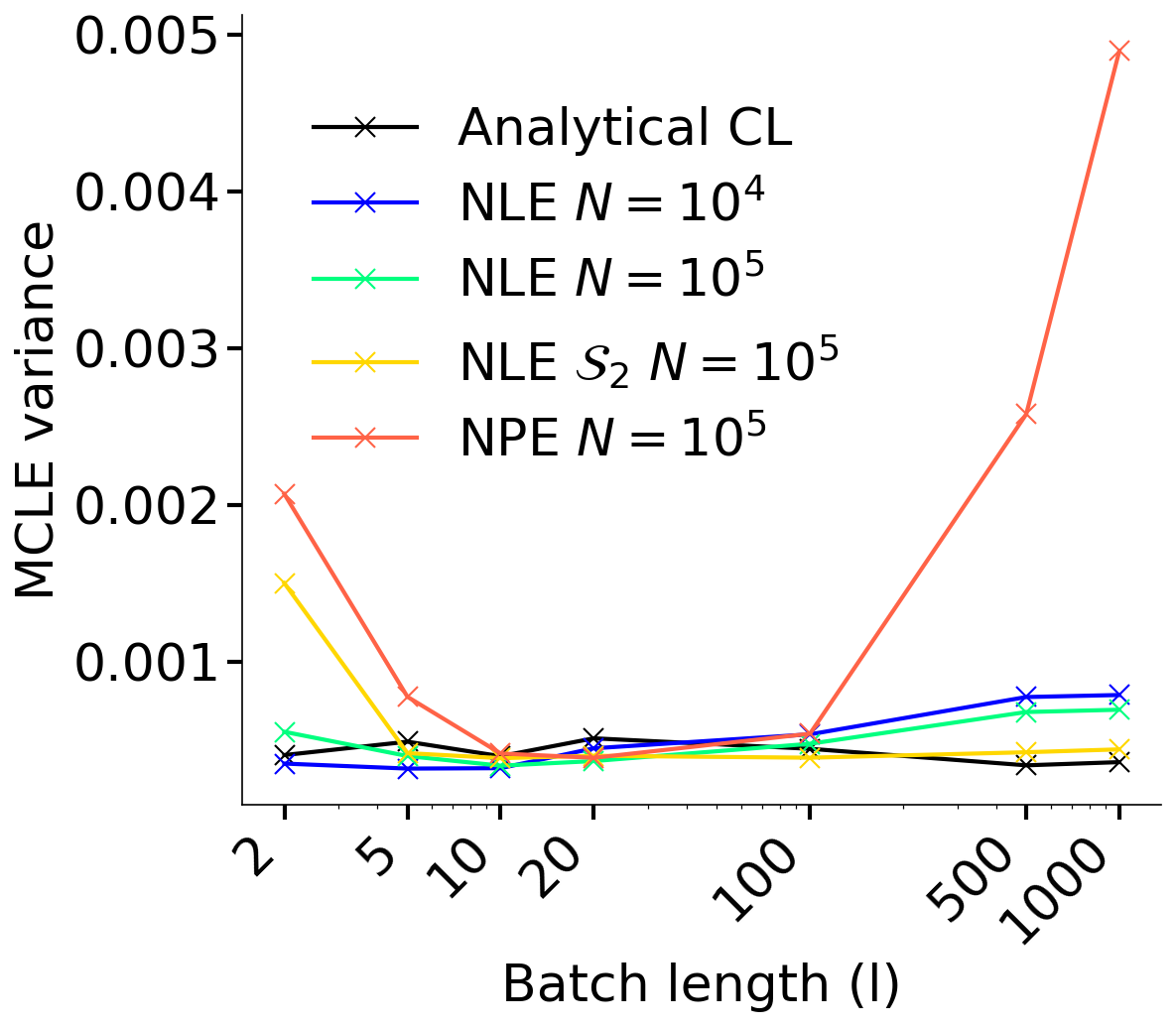}
            \put(-140,125){\textbf{(h)}}
            \phantomcaption
            \label{fig:ar1_mcle_mean}
        \end{subfigure}
    \end{subfigure}
    \hfill
    \begin{subfigure}[t]{0.35\textwidth}
        \centering
        \begin{subfigure}{\textwidth}
            \centering
            \includegraphics[width=0.9\linewidth]{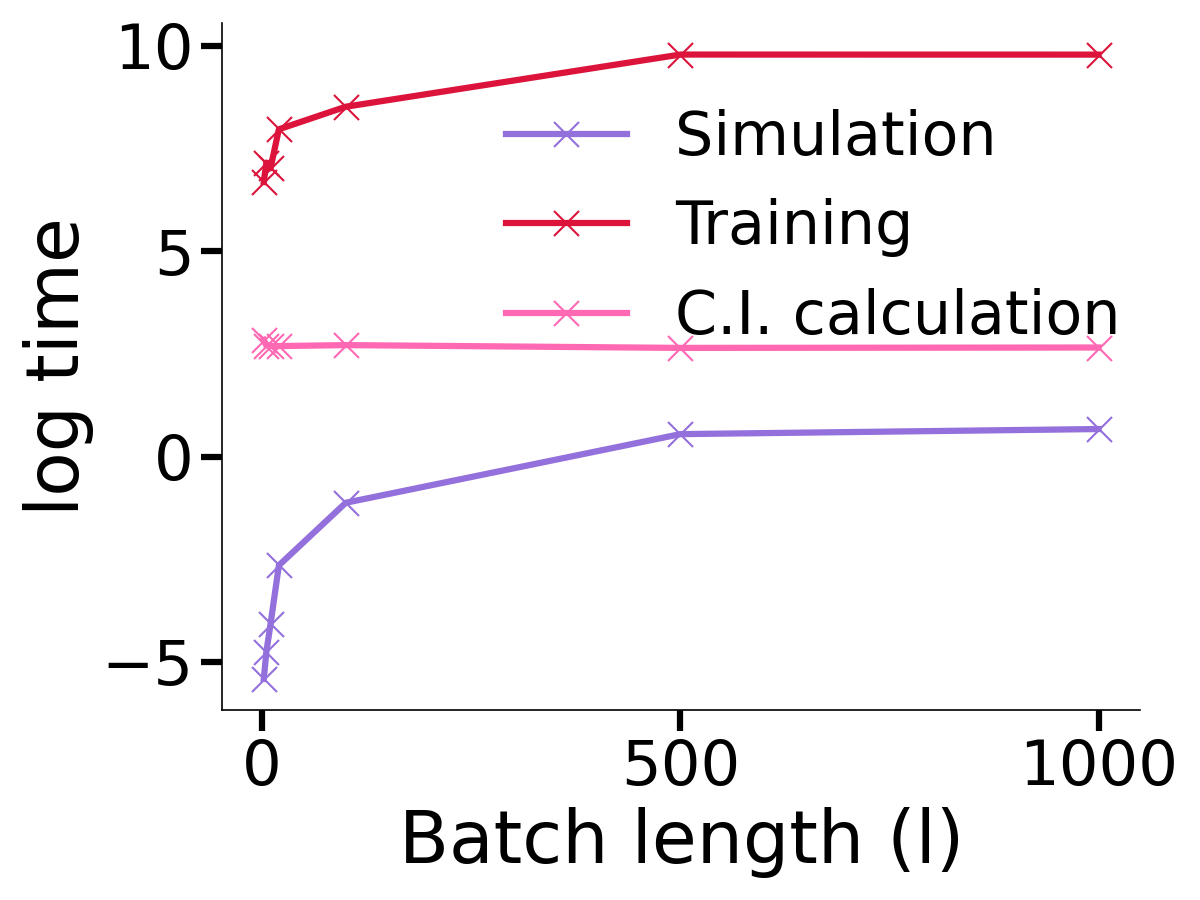}
            \put(-150,115){\textbf{(f)}}
            \phantomcaption
            \label{fig:ar1_mcle_variance}
        \end{subfigure}
        \begin{subfigure}{\textwidth}
            \raisebox{3cm}{\includegraphics[width=1\linewidth]{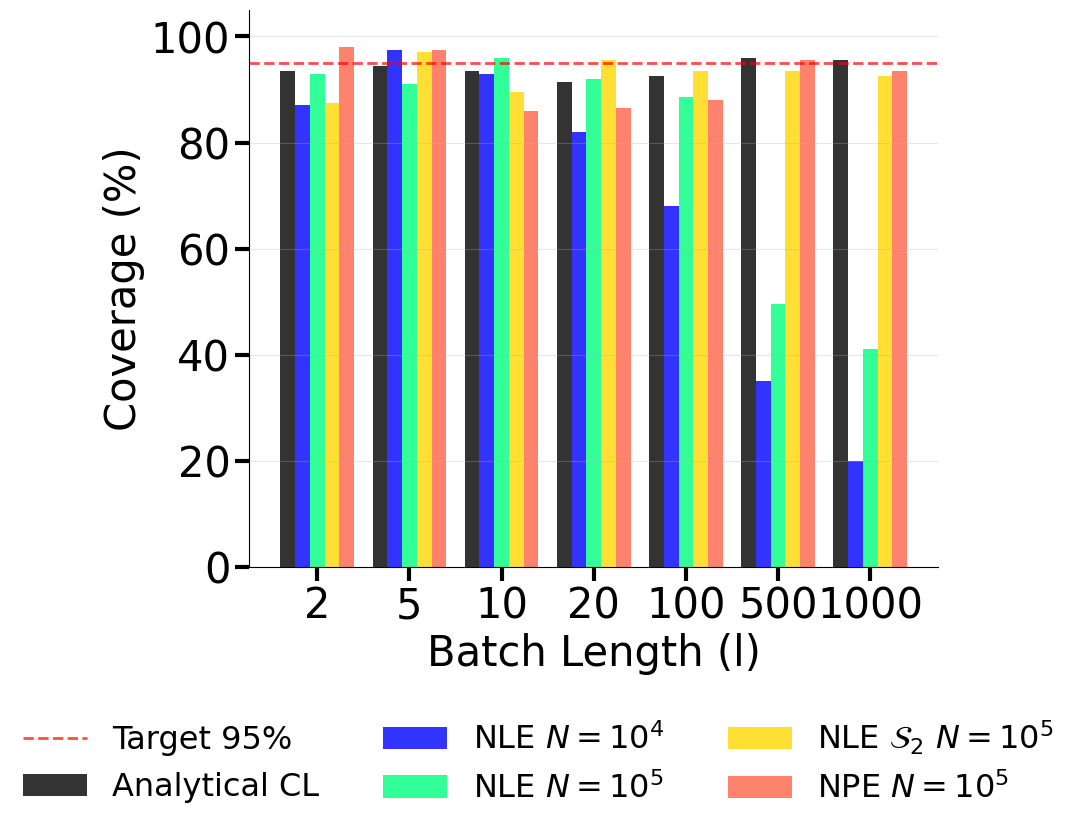}}
            \put(-155,200){\textbf{(i)}}
            \phantomcaption
            \label{fig:ar1_coverages}
        \end{subfigure}
    \end{subfigure}
    \vspace{-2cm}
    \caption{AR(1) model. Normalised NCLs trained on $N=10^5$ samples using (a) raw data and (b) sufficient statistics $\mathcal{S}_2$. (c) Normalised analytical CLs. (d) RMSE between analytical CL and raw data NCLE log-likelihoods. (e) Confidence interval widths. (f) Mean log computation times for raw data NLE (in seconds). (g) Mean of 200 MCLEs. (h) Variance of 200 MCLEs. (i) 95\% coverages.}
    \label{fig:ar1_plots}
\end{figure}

Figure \ref{fig:ar1_plots} presents our NCLE results for the AR(1) model. We describe each plot in turn.
Some plots are aggregated results over multiple test datasets.

Figures \ref{fig:ar1_ncls} and \ref{fig:sufficientSS_ncls} show the normalised likelihood \eqref{eq:normalised_ncl} estimated by the NCL trained on \(N=10^5\) training samples
using the raw data and the sufficient statistics $\mathcal{S}_2$ defined in \eqref{eq:ar1_sufficientSS}. Figure \ref{fig:ar1_cls} shows the normalised likelihood estimated by the analytical CL using the raw data.
For analytical CL, all the curves have a similar mode to the true likelihood, but they become slightly narrower for $l=2,10$.
This agrees with Appendix \ref{sec:ar1_deriv} which shows the composite likelihood for AR(1) has higher asymptotic curvature (and so is likely over-concentrated) when $|\theta|>\frac{1}{\sqrt{3}}$, especially for smaller $l$.
In the figure, $l=100$ gives a composite likelihood approximation with negligible visual error to the likelihood shape.
The NCLE curves are similar, although some curves -- $l=100$ using raw data and $l=2$ using $\mathcal{S}_2$ -- have their mode further away from the MLE, indicating some NLE approximation error.
(Reasons for the poor $\mathcal{S}_2$ result are discussed below.)

Figure \ref{fig:ar1_rmses} shows the RMSE \eqref{eq:rmse} between the (raw data) NCL and analytical CL as $l$ varies.
The RMSE values are substantially larger for $l>100$.

Figure \ref{fig:ar1_ci_widths} shows mean confidence interval width for NCLE and analytical CL as $l$ varies, with each average computed from 200 confidence intervals.
Table \ref{tab:ar1_ci_widths} in the appendix gives more numerical details.
Analytical CL confidence interval widths are mostly unaffected by batch length.
NCLE widths vary much more, illustrating substantial NLE approximation error.
Increasing the raw data NLE training sample size from $N=10^4$ to \(10^5\) gives NCLE widths roughly matching analytical CL results for $l \leq 100$.
NLE with $\mathcal{S}_2$ gives generally accurate widths for any $l>2$.
NPE closely matched analytical CL results only for $l \in [10,100]$.

Figure \ref{fig:ar1_times} shows log computation times plotted against batch length for NCLE with \(N=10^5\) using raw data,
with more numerical detail in Table \ref{tab:ar1_times}.
For each \(l\), the sampling and training times are averaged over 10 runs, while confidence interval times are averaged over 200 runs.
The total time required to generate and train on simulated data decreases substantially as $l$ is reduced. 
Sampling the training data takes very little time for any $l$: less than two seconds.
The mean time required to compute the confidence intervals is roughly constant.
Thus, computational cost differences across \(l\) are mainly due to neural network training.

Figures \ref{fig:ar1_mcle_mean} and \ref{fig:ar1_mcle_variance} show the sample mean and variance of 200 MCLEs respectively, for NCLE and analytical CL as $l$ varies. 
Finally, Figure \ref{fig:ar1_coverages} shows 95\% coverage computed from 200 confidence intervals for NCLE and analytical CL, as $l$ varies.
Table \ref{tab:ar1_coverages} in the appendix gives more numerical detail on coverage.
We comment on these figures in the next subsection 
as they are particularly important for our conclusions.


\subsubsection{Comments} \label{sec:AR1comments}

Overall, this example illustrates that NCLE can perform better than standard SBI (i.e.~NPE or NLE with $l=L=1000$) in terms of inference quality and computational cost.
Below we discuss other findings.

Analytical composite likelihood produces good estimates for any choice of $l$.
This verifies empirically that the CL asymptotics perform well, despite lacking full theoretical support.
NCLE methods often have worse performance.
This illustrates that the SBI approximation used in NCLE can introduce errors,
and that tuning choices are crucial to controlling the amount of error and the computational cost.

The results reflect several tuning trade-offs described in Section \ref{sec:batch_length}.
First consider NLE using raw data.
Large $l$ produces poor coverage, reflecting difficulty estimating high dimensional $\mathbf{x}_b$.
Coverage was especially poor for $l \geq 100$ with $N=10^4$ training samples, or $l \geq 500$ with $N=10^5$.
Increasing $N$ from $10^4$ to $10^5$ also improves performance across other diagnostics, such as MCLE bias.
We found computation costs increased with large $l$, driven by higher costs in training rather than data simulation.

Likelihood estimation using NPE on raw data also has issues.
Large $l$ produces high variance MCLEs (Figure \ref{fig:ar1_mcle_variance})
and over-wide CIs (Figure \ref{fig:ar1_ci_widths}).
We speculate that NPE struggles with data summarisation for high dimensional $\mathbf{x}_b$
-- as discussed in Section \ref{sec:batch_length} --
effectively producing a poor summary $\mathbf{s}(\mathbf{x}_b)$ which loses much information
(i.e.~this is produced part-way through the NPE neural network).
However, the resulting coverage is good (Figure \ref{fig:ar1_coverages}).
We speculate this is because NPE performs accurate density estimation of $p(\mathbf{s}|\bm{\theta})$.

Under NPE, small $l$ produces biased MCLEs and over-wide CIs.
The reason may be difficulty estimating the posterior for small $l$: exploratory plots show the posterior approximation often has mass outside the support $[-1,1]$.
NPE works best for a middle range of $l$ values, roughly $[10,100]$, as the CI width and MCLE variance both increase for $l$ values outside this range.

Overall, NLE has good performance under a wider range of $l$ than NPE.
Therefore the next section focuses on NLE.

NCLE using the sufficient statistics $\mathcal{S}_2$ performs well for all batch lengths except $l=2$,
where performance is poor.
The reason is that our code implementation makes one statistic a constant (matching the derivation in Appendix \ref{sec:ar1_deriv}),
a situation in which NLE performs poorly (under default tuning choices).
For $l>2$, these summaries avoid problems with learning to summarise data, or high dimensional density estimation.
So in practice, one could use $\mathcal{S}_2$ with $l=5$ or $10$ to get good inference results.
This motivates using fixed length summaries in other applications
even when low dimensional sufficient statistics are not available.

It's reassuring that all methods have good coverage for some $l$ values.
This provides empirical support for using the confidence interval methods from Section \ref{sec:ci},
despite the various approximations involved in their derivation
(e.g.~reliance on Bartlett identities).

Section \ref{sec:batch_length} suggested using a simulation study to tune $l$.
As an illustration, consider NLE with raw data.
The results above support using $l=10$,
as it gives good MCLE accuracy and reasonable coverage.
This is a good match to the heuristic choice from Section \ref{sec:ar1_batch_length}.
(The empirical ACF suggested a larger value, $l \approx 20$, which also has reasonable performance in practice.)
Using Table \ref{tab:ar1_times}, we can see that inference using $l=10$ is faster than $l=1000$ by a factor of roughly 16.

\subsection{GARCH model} \label{sec:GARCH}

We use the following GARCH(1,1) model \citep{bollerslev1986garch},
\begin{align}
    &x_t \sim N(\mu, \sigma^2_t), \label{eq:GARCHx} \\
    &\sigma^2_t = \omega + \alpha (x_{t-1}-\mu)^2 + \beta \sigma^2_{t-1}. \label{eq:GARCHsigma} 
\end{align}
We want to use the stationary distribution,
where $t \in \mathbb{Z}$ and we make observations at $t=1,\ldots,L$.
However, simulating from this is intractable.
Therefore we approximate it using burn-in,
by sampling $500+L$ steps given
$\sigma_1^2 = \frac{\omega}{1-\alpha-\beta}$,
and taking the final $L$ outputs.
Using 500 burn-in steps matches the \texttt{arch} package \citep{arch}.
As further justification we note that \citet{li2024burn} report that
in experiments with GARCH(1,1) models the number of burn-in steps is not critical,
and even a small number of steps can achieve good performance.
(Indeed we note in Section \ref{sec:long_seq} that, for one particular experiment,
not using burn-in also performs well.)

Our parameters of interest are $\bm{\theta}=(\mu, \omega, \alpha, \beta)$,
with constraints $\alpha \geq 0, \beta \geq 0, \omega > 0$.
We use a further constraint for stationarity, $\alpha+\beta<1$.
Our synthetic observations are simulated under $\bm{\theta}^*=(0.5, 0.1,0.1, 0.8)$.

The likelihood under stationarity is not tractable.
As a baseline comparison, we use conditional maximum likelihood estimation (CMLE) \citep{francq2019garch}, which approximates the stationary likelihood by optimizing:
\begin{equation} \label{eq:garch_conditional_likelihood}
p(\mathbf{x}|\mu, \omega, \alpha, \beta) = 
\Phi(x_1; \mu, \sigma_1^2)
\prod_{t=2}^L \Phi \left( x_t; \mu, \sigma_t^2 \right),
\end{equation}
and fixes $\sigma_1^2 = \frac{\omega}{1-\alpha-\beta}$.
This objective involves the values $\sigma_2, \sigma_3, \ldots, \sigma_L$,
which are a deterministic transformation of $\mathbf{x}$ by applying \eqref{eq:GARCHsigma} recursively.
For sufficiently large $L$, CMLE is a good estimate of the MLE under stationarity.
For an analytical composite likelihood baseline,
we use composite likelihood based on \eqref{eq:garch_conditional_likelihood}.
(In this case, an approximation to the desired stationary likelihood term is used for the start of each batch,
which could produce significant error when $l$ is small.)

Appendix \ref{app:regularity} shows that, as for the AR(1) model, the existing composite likelihood regularity conditions aren't met and again we must verify asymptotic behaviour empirically.

Initially, in Sections \ref{sec:GARCH_results}--\ref{sec:garch_comments}, we investigate sequence length $L=1000$
to select a batch length $l \in \{2,5,10,20,100,500,1000\}$,
the same set as for AR(1).
Then, in Section \ref{sec:long_seq}, we use our optimal $l$
for a longer sequence, of length $L=10^6$.
Taking $l=L$ corresponds to ordinary NLE.
We consider training data size $N \in \{10^4, 10^5,10^6\}$.
For this example we work with raw data i.e.~$S(\mathbf{x}_1^{(i)})=\mathbf{x}_1^{(i)}$.
However we perform NLE under a reparameterisation, described below.

\subsubsection{Reparameterisation}

We found the parameters $\omega$, $\alpha$ and $\beta$ can be difficult to identify.
We therefore trained the neural network to learn the likelihood of the reparameterisation $\bm{\psi} = ( \mu, \varsigma^2, \gamma, \delta )$ instead,
where $\gamma \coloneq \alpha+\beta$ (known as the \textit{persistence}),
$\varsigma^2 \coloneq \omega/(1-\gamma)$ and $\delta \coloneq \frac{\alpha}{\gamma}$.
Another advantage of using these new parameters is we found they made the calculation of the Godambe matrix more numerically stable.
We used the following training distributions:
\[ \mu \sim U(-1,1), \quad \varsigma^2 \sim U(0,2), \quad \gamma \sim U(0,1), \quad \delta \sim U(0,1). \]
These distributions of $\gamma$ and $\delta$ meet the required constraints on $\alpha$ and $\beta$.

The MCLE for $\bm{\theta}$ was recovered from the MCLE for $(\mu, \varsigma^2, \gamma, \delta)$ using
$\omega = \varsigma^2(1-\gamma)$, $\alpha = \gamma\delta$,  $\beta = \gamma(1-\delta)$.
The Godambe matrix for $\bm{\theta}$ is $\mathbf{G}_{\bm{\theta}} = J^\top \mathbf{G}_{\bm{\psi}} J$,
where $\mathbf{G}_{\bm{\psi}}$ is the Godambe matrix for $\bm{\psi}$, and
$J$ is the Jacobian of the transformation $\bm{\theta} \mapsto \bm{\psi}$.
So we estimated $\mathbf{G}_{\bm{\psi}}$ using Algorithm \ref{alg:gim},
and then transformed it using $J$.
We used the resulting estimate of $\mathbf{G}_{\bm{\theta}}$ to compute confidence intervals.

\subsubsection{Batch length heuristic} \label{sec:GARCH_batch_length}

For GARCH a typical measure of autocorrelation \citep{francq2019garch} is:
\begin{equation*}
C(k) = \mathrm{Corr}\left((x_t-\mu)^2,(x_{t-k}-\mu)^2\right)
= \frac{\alpha(1 - \beta(\alpha + \beta))}{1 - (\alpha + \beta)^2 + \alpha^2} (\alpha+\beta)^{k-1},
\end{equation*}
for $k \geq 1$.
We find numerically that, under $\bm{\theta}^*$, $C(k) < c$ with $c=0.1$ when $k \geq 5$, suggesting that \(l \ge 5\) is sufficient to capture short-range dependence in the data.
We also calculated the empirical autocorrelation using synthetic data $\mathbf{x}^{obs}$ (of length $L=1000$).
This dropped below $c$ at the small value $k=2$. Our results investigate how well these values behave as batch lengths in practice.

\subsubsection{Results} \label{sec:GARCH_results}

A small number of cases produced numerical errors, so we removed these from our results.
Occasional replications (about 1 in 200) failed,
for data simulated from $\delta \approx 1$ i.e.~persistence close to the bound of stationarity.
Batch lengths \(l=2,5\) are omitted from some plots due to ill-conditioning of the Godambe matrix for the analytical CL.
NCLE methods did not experience this problem,
possibly because NLE approximation error acts to smooth out the ill-conditioning,
or because they approximate the stationary likelihood rather than \eqref{eq:garch_conditional_likelihood}.

\begin{figure}[tbp]
    \centering
    \begin{subfigure}{0.49\textwidth}
        \centering
    \includegraphics[width=1\linewidth]{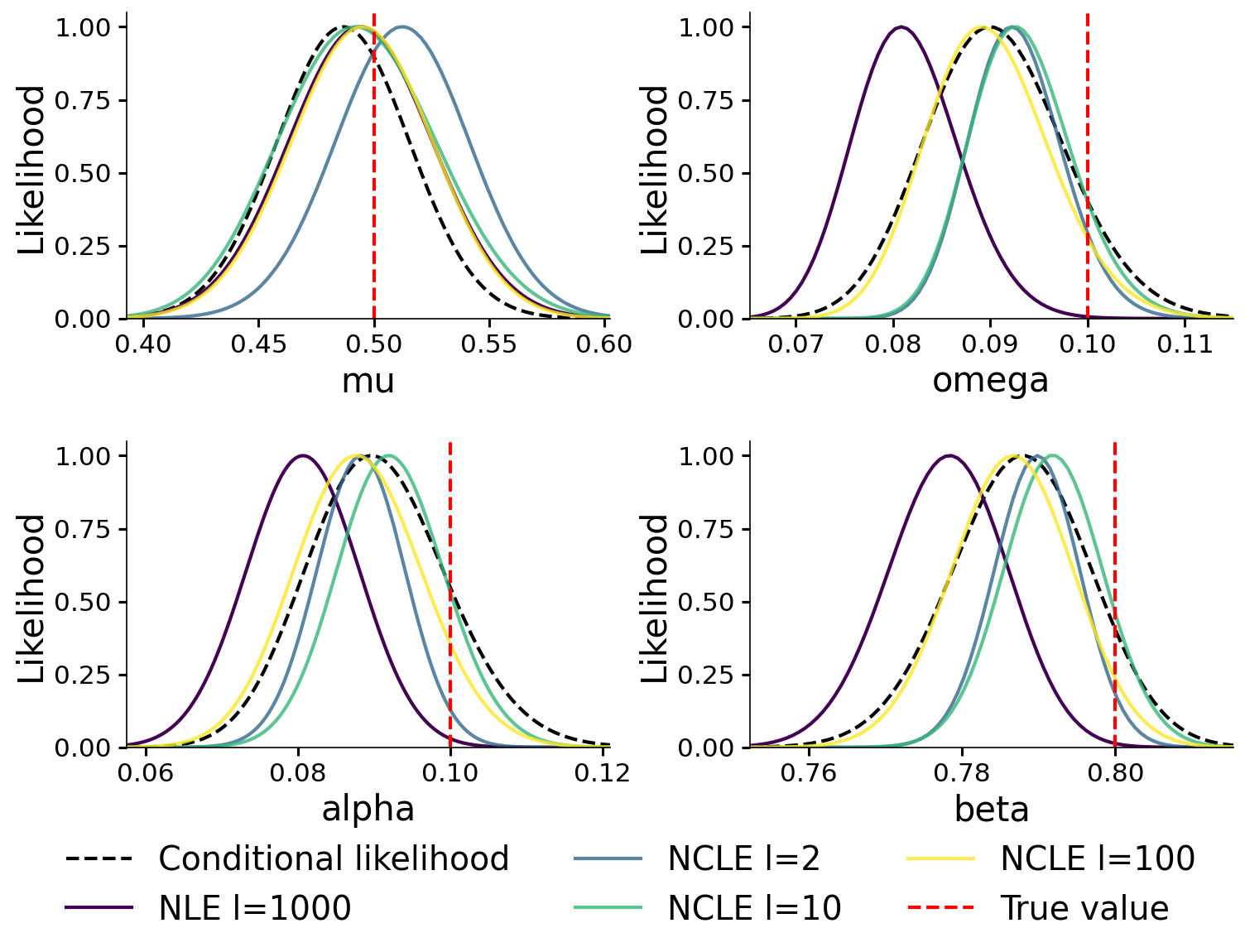}
    \put(-240,170){\textbf{(a)}}
    \phantomcaption  
    \label{fig:garch_ncls}
\end{subfigure}
\hfill
\begin{subfigure}{0.49\textwidth}
    \centering
    \includegraphics[width=1\linewidth]{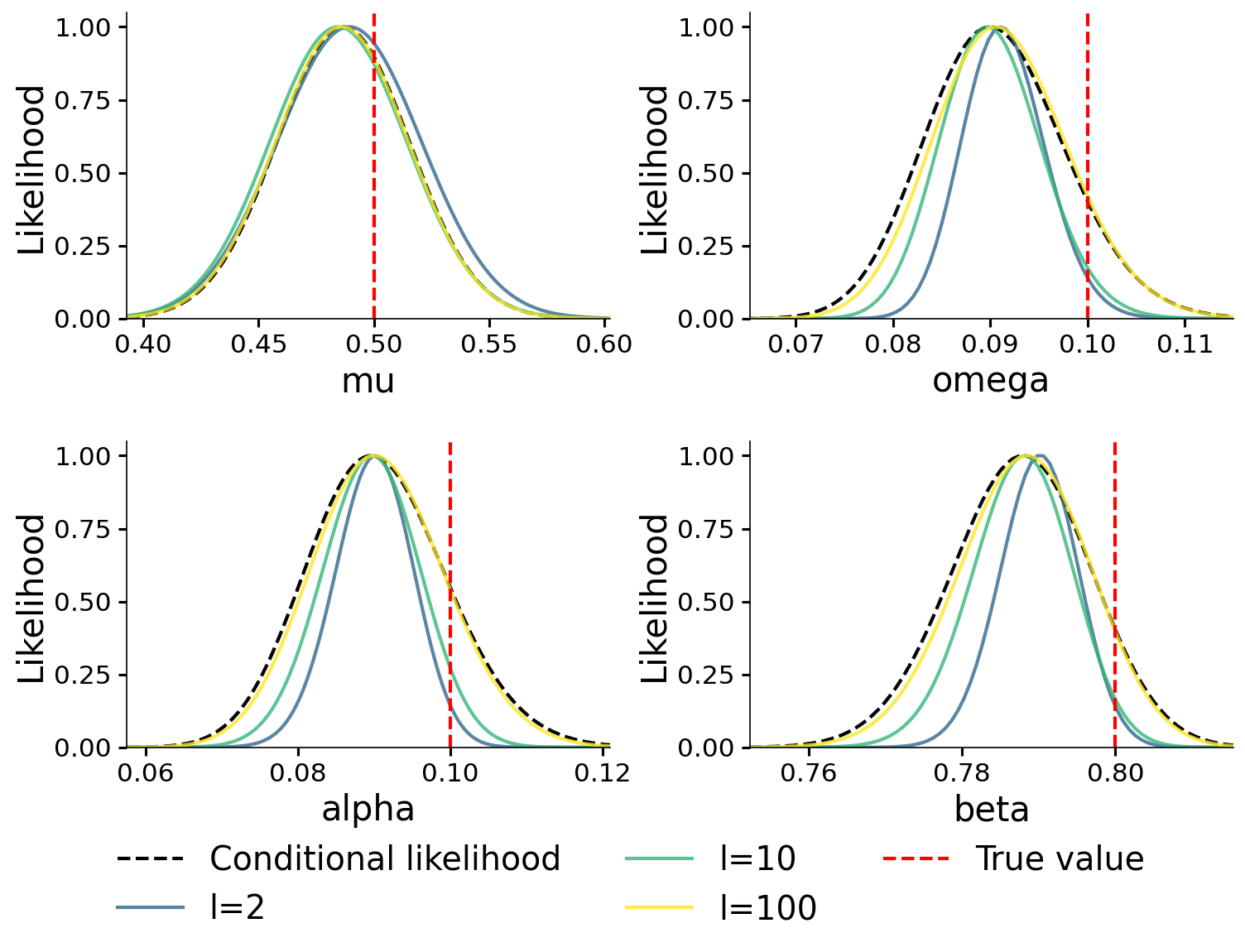}
    \put(-240,170){\textbf{(b)}}
    \phantomcaption  
    \label{fig:garch_cls}
\end{subfigure}
\hfill
\begin{subfigure}{0.4\textwidth}
    \centering
    \includegraphics[width=1\linewidth]{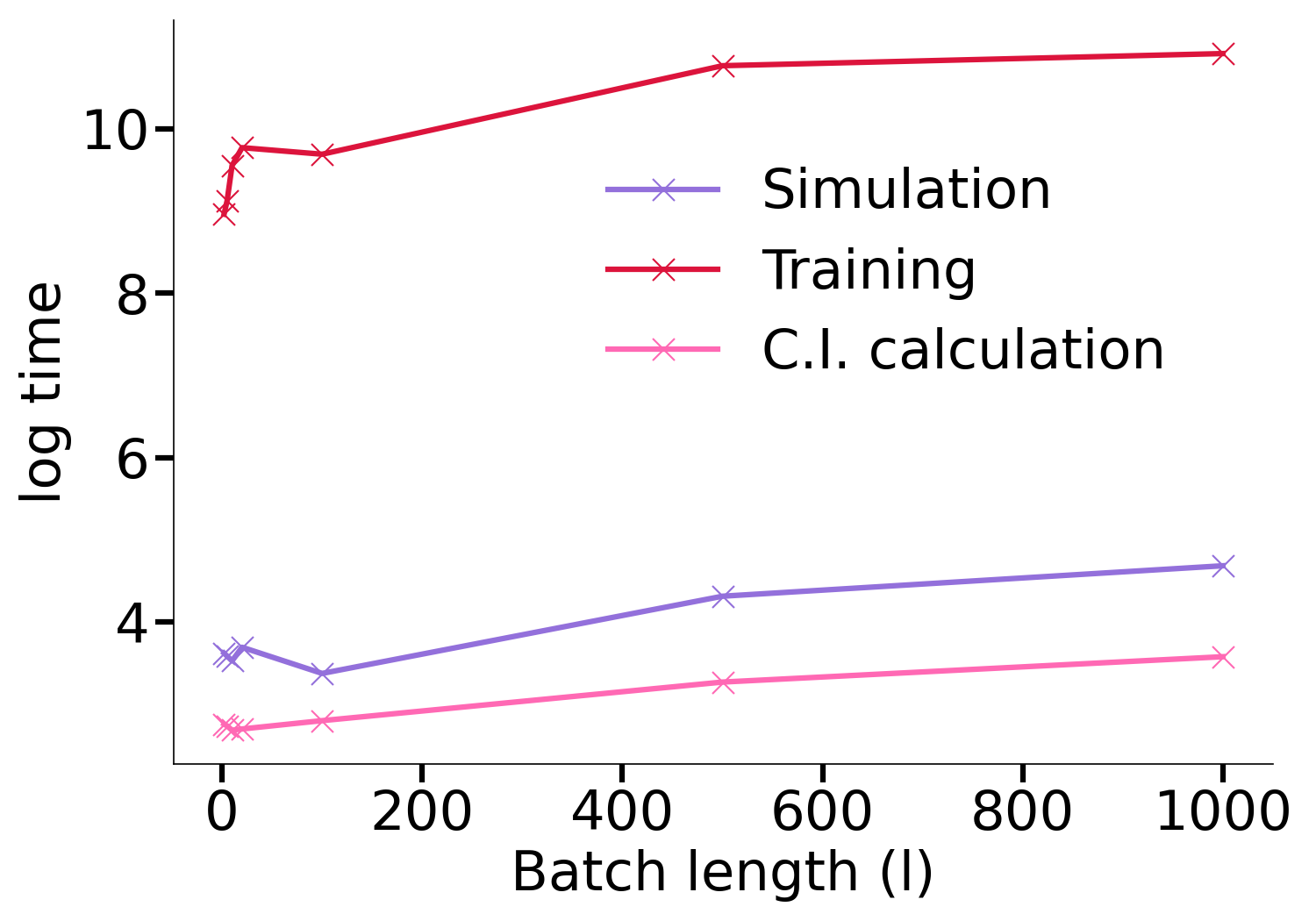}
    \put(-185,120){\textbf{(c)}}
    \phantomcaption  
    \label{fig:garch_times}
\end{subfigure}
\hspace{0.02\textwidth}
\begin{subfigure}{0.4\textwidth}
        \centering
    \includegraphics[width=1\linewidth]{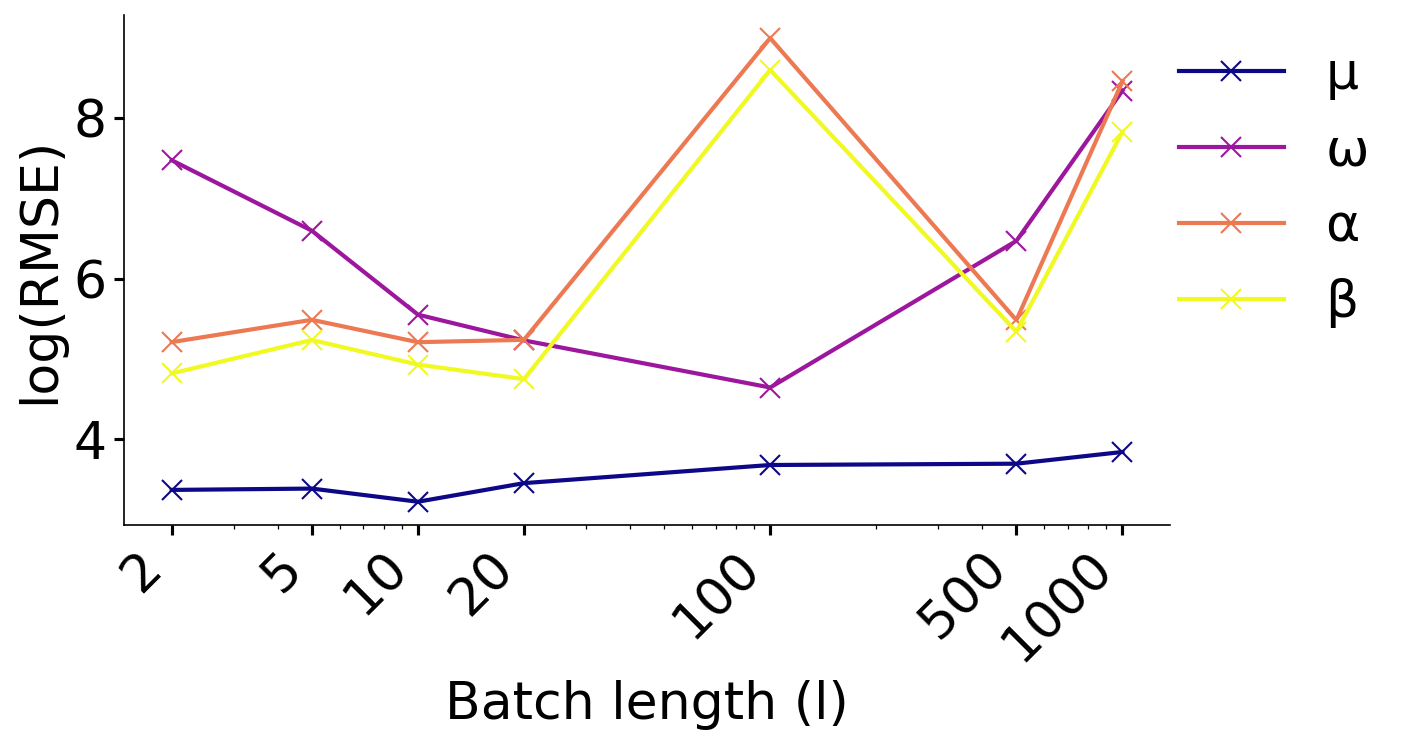}
    \put(-180,120){\textbf{(d)}}
    \phantomcaption  
    \label{fig:garch_rmses}
\end{subfigure}

    \caption{GARCH(1,1) model. (a) Normalised NCLs trained on $N=10^6$ samples.
    (b) Normalised analytical CLs.
    (c) Mean log computation times using $N=10^6$ (in seconds).
    (d) log(RMSE) between CL and NCL log-likelihoods using $N=10^6$. }
    \label{fig:garch_plots}
\end{figure}

\begin{figure}[tbp]
    \centering
    \begin{subfigure}{0.9\textwidth}
        \centering
    \includegraphics[width=1\linewidth]{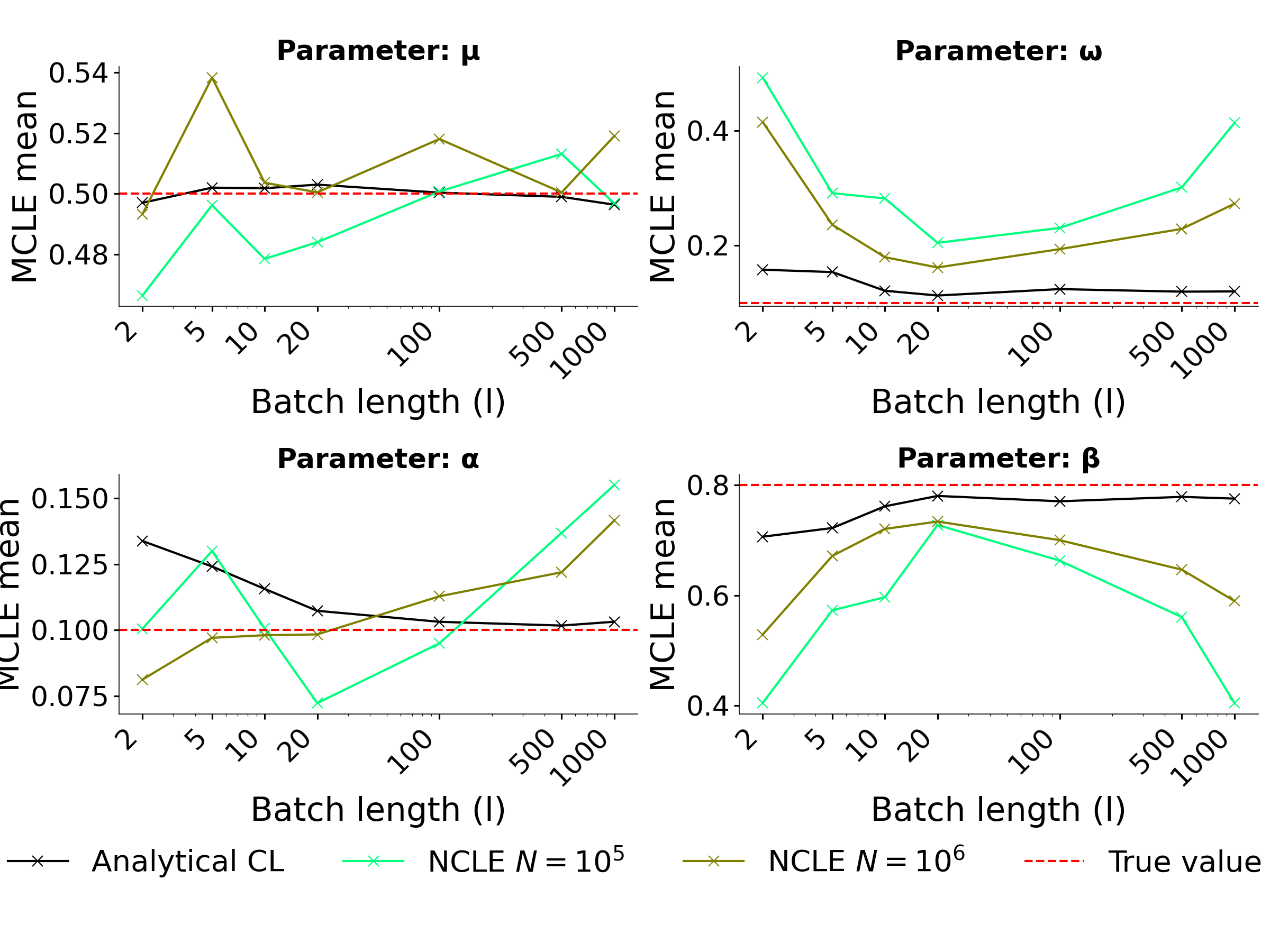}
    \put(-415,300){\textbf{(a)}}
    \phantomcaption  
    \label{fig:garch_mcle_means}
\end{subfigure}\\[1em]
    \begin{subfigure}{0.9\textwidth}
        \centering
    \includegraphics[width=1\linewidth]{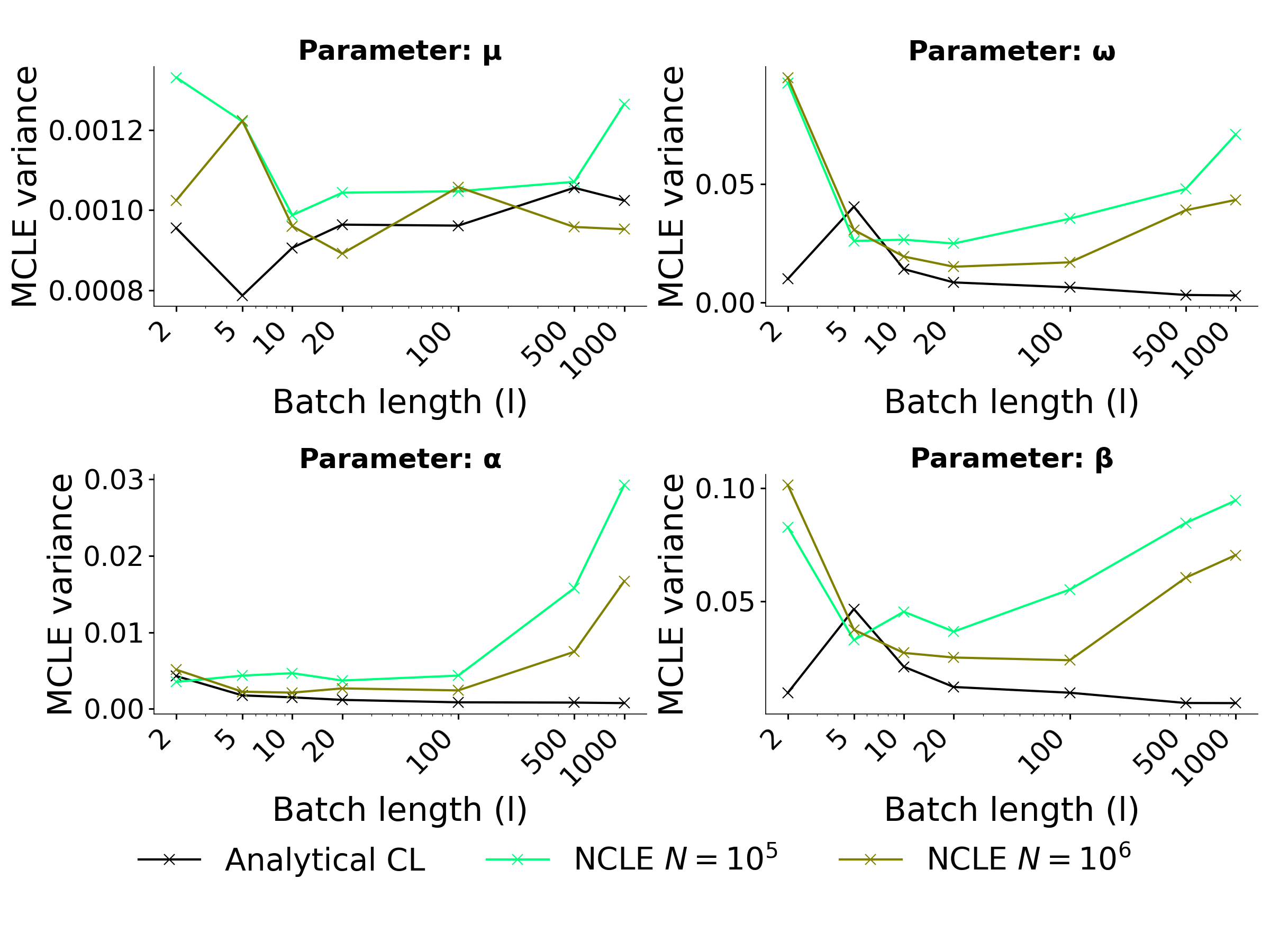}
    \put(-425,300){\textbf{(b)}}
    \phantomcaption  
    \label{fig:garch_mcle_variances}
\end{subfigure}
\vspace{-5pt}
\caption{GARCH(1,1) model. (a) Mean of 200 MCLEs. (b) Variance of 200 MCLEs.}
\label{fig:garch_plots2}
\end{figure}

\begin{figure}[tbp]
    \centering
    \begin{subfigure}{0.65\textwidth}
        \centering
        \includegraphics[width=1\linewidth]{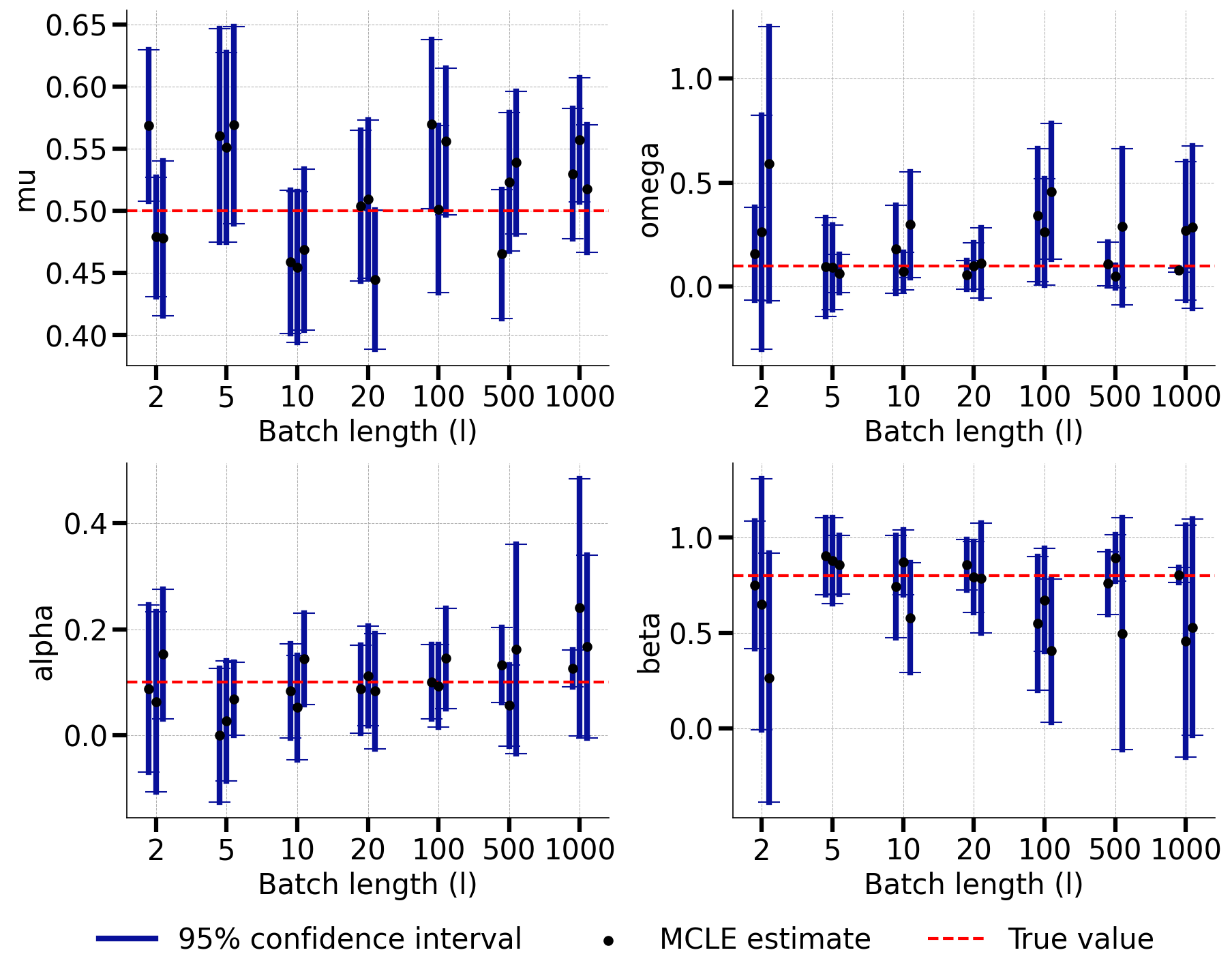}
        \put(-295,235){\textbf{(a)}}
        \phantomcaption
        \label{fig:garch_cis}
    \end{subfigure}
    \hfill
    \begin{subfigure}{0.34\textwidth}
        \centering
        \begin{subfigure}{\textwidth}
            \centering
            \includegraphics[width=1\linewidth]{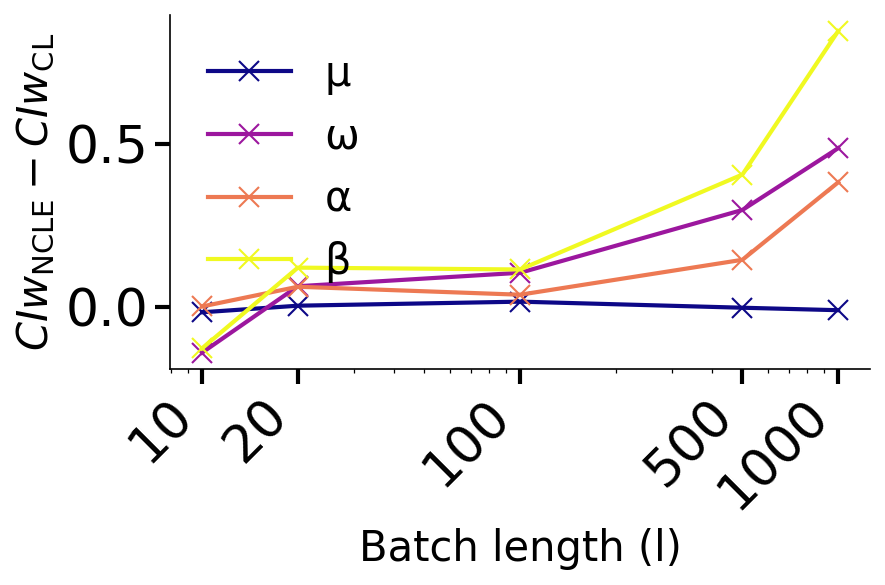}
            \put(-155,105){\textbf{(b)}}
            \phantomcaption
            \label{fig:garch_ci_widths_diffs}
        \end{subfigure}

        \begin{subfigure}{\textwidth}
            \centering
            \includegraphics[width=1\linewidth]{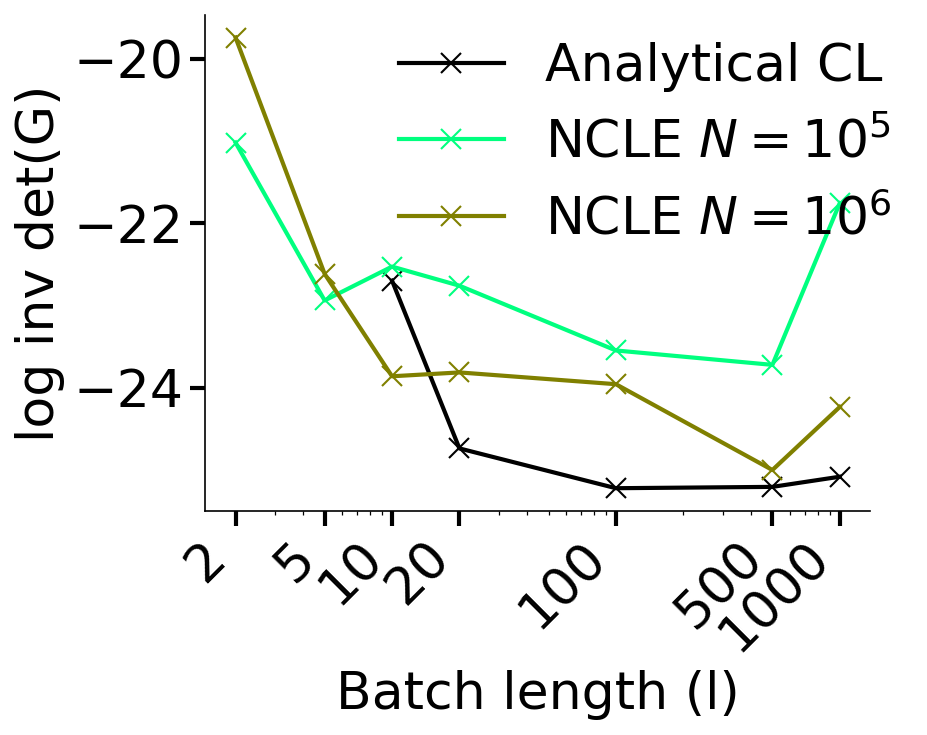}
            \put(-155,120){\textbf{(c)}}
            \phantomcaption
            \label{fig:garch_det}
        \end{subfigure}
    \end{subfigure}\\[0.5em]

    \begin{subfigure}{0.75\textwidth}
        \centering
    \includegraphics[width=1\linewidth]{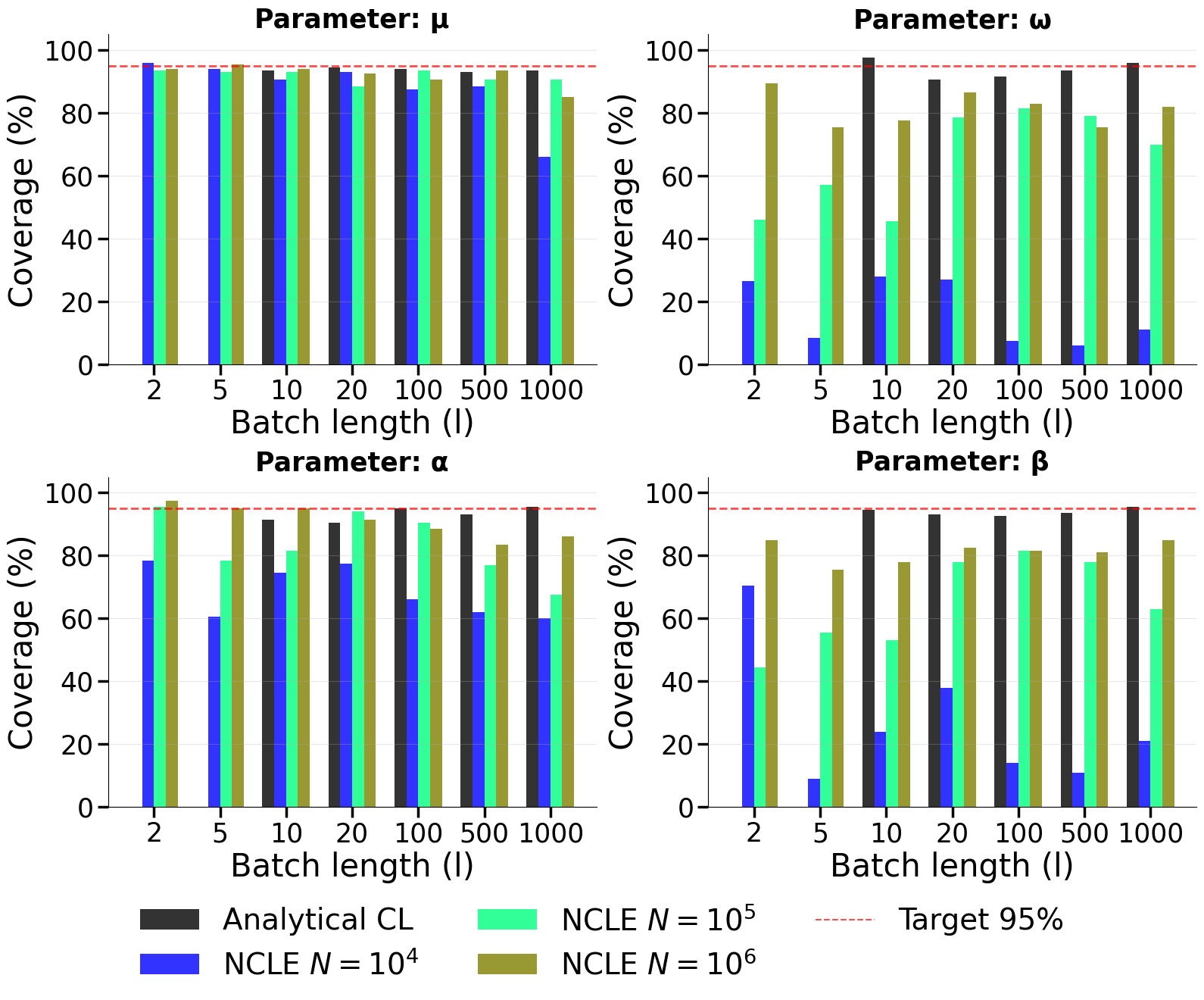}
    \put(-340,280){\textbf{(d)}}
    \phantomcaption  
    \label{fig:garch_coverages}
\end{subfigure}
    \caption{GARCH(1,1) model. (a) NCLE CIs for $N=10^6$. (b) $CIw_\mathrm{NCLE} - CIw_\mathrm{CL}$ for $N=10^6$. Larger positive values mean NCLE has wider CIs than CL. (c) Mean $\log\det(\mathbf{G}^{-1})$ calculated from 200 CIs. More negative values indicate greater precision. (d) 95\% coverages.
    Analytical CL results are omitted for $l=2,5$ due to numerical problems in calculation.
    }
    \label{fig:garch_plots3}
\end{figure}

Figures \ref{fig:garch_plots}--\ref{fig:garch_plots3} present our NCLE results for the GARCH(1,1) model.
We describe each plot in turn.

Figure \ref{fig:garch_ncls} shows the normalised composite likelihood \eqref{eq:normalised_ncl} for each parameter using
NCL with \(N=10^6\).
Figure \ref{fig:garch_cls} instead uses the analytical CL. 
All plots are for the same synthetic data $\mathbf{x}^{obs}$.
In both plots, the curves vary more with $l$ than for the AR(1) results, illustrating the importance of this tuning choice.
The NCLE curves have more variation,
likely reflecting NLE approximation error
(although the analytic CL curves also have errors from using approximate likelihood terms based on \eqref{eq:garch_conditional_likelihood}).
The NCLE curves for $l=1000$ are particularly poor,
likely reflecting poor NLE performance estimating a high dimensional density.

Figure \ref{fig:garch_times} shows log computation times plotted against batch length using NCLE with \(N=10^6\),
with more details in Table \ref{tab:garch_times}.
For each \(l\), the sampling and training times were averaged over 5 runs, while confidence interval times were averaged over 200 runs.
The time required to generate and train on simulated data decreases substantially for shorter batches.
For instance, generating training samples with \(l=10\) was approximately 3 times faster than using full-length sequences
(which is consistent with linear scaling of simulation time with total number of steps, including burn-in),
while neural network training was almost 4 times faster.
Unlike the AR(1) experiment, CI calculation time clearly increases with $l$ overall
(for $l$ large enough i.e.~$l \geq 100$).
For example, using $l=10$ was roughly $2.5$ times faster than the full-length sequence.

Figure \ref{fig:garch_rmses} shows the log RMSE, from  \eqref{eq:rmse}, between the NCL with \(N=10^6\) and the analytical CL, as $l$ varies.
The RMSE for \(\mu\) was the smallest among the parameters, reflecting that this mean parameter is easy to learn.

Figures \ref{fig:garch_mcle_means} and \ref{fig:garch_mcle_variances} show the sample mean and variance of 200 MCLEs respectively, plotted for NCLE and analytical CL as $l$ varies.
Increasing $N$ generally improves the NCLE mean and variance, although for some parameters there is little difference
(e.g.~$\alpha$ for mean and $\mu$ for mean and variance). The MCLE variance for \(\mu\) was approximately an order of magnitude smaller than for the other parameters.
Overall, there is no consistent pattern across parameters of which batch lengths yielded MCLEs closest to the true value or the lowest variance.

Figure \ref{fig:garch_cis} shows three confidence intervals per batch length for each parameter using NCLE with \(N=10^6\).
The typical confidence intervals for \(\mu\) were narrower than those for the other parameters,
while $\omega$ and \(\beta\) generally had the widest intervals.
Most confidence intervals contain the true parameter, but there is considerable variation in their length.
Indeed several intervals contain negative values, which are outside the parameter support.

Figure \ref{fig:garch_ci_widths_diffs} compares NCLE (\(N=10^6\)) and analytical CL confidence interval widths by plotting \(CIw_{\mathrm{NCLE}} - CIw_{\mathrm{CL}}\) as $l$ varies.
Here \(CIw\) denotes the mean confidence interval width computed from 200 intervals.
Table \ref{tab:garchCIwidths} has more details.
Larger positive values indicate wider NCLE intervals relative to analytical CL. For \(\mu\), the difference was small for all \(l\), whereas for other parameters the difference increases with \(l\).
This suggests that NLE approximation error increases with $l$ (at least for $l \geq 10$).

Figure \ref{fig:garch_det} shows mean \(\log \det(\mathbf{G}^{-1})\) from 200 datasets using NCLE and analytical CL, as $l$ varies.
See also Table \ref{tab:garch_det} for more numerical detail.
Smaller (more negative) values indicate Godambe matrices with greater precision,
and correspond to asymptotic confidence ellipsoids with smaller size.
NCLE with more training samples generally gave smaller values for $l \geq 10$,
which were typically closer to the analytical values.
The \(\log \det(\mathbf{G}^{-1})\) values generally decrease initially as $l$ grows,
but eventually plateau or even increase for larger $l$.
However the best choice of $l$ differs between the lines shown.

Finally, Figure \ref{fig:garch_coverages} shows a bar plot of 95\% coverage computed from 200 confidence intervals using NCLE and analytical CL, plotted against batch length.
Table \ref{tab:garch_coverages_combined} gives more numerical details.
Analytical CL gives reasonable coverage for all choices of $l \geq 10$.
For $l<10$ we could not compute this coverage due to numerical problems (described at the start of this section).
NCLE coverage quality is often worse, which we discuss further in the next section.





\subsubsection{Comments} \label{sec:garch_comments}

As for the AR(1) model, this example illustrates that NCLE can perform better than standard SBI
(i.e.~NLE with a single batch, corresponding to $l=1000$)
in terms of inference quality and computational cost.
Below we discuss other findings.

For $l$ sufficiently large, analytical composite likelihood performs similarly to using an analytical likelihood
(i.e.~the $l=1000$ case).
For instance the MCLE means and variances in Figure \ref{fig:garch_plots2} are similar for any $l \geq 100$.
This suggests that the CL asymptotic results \eqref{eq:asymptoticMCLE} hold reasonably well.
Worse performance for smaller $l$ may be partially due to error from using \eqref{eq:garch_conditional_likelihood} rather than the stationary likelihood.
So we shouldn't rule out using $l<100$ with NCLE.


NCLE methods often have worse performance than analytical CL.
This illustrates again that the SBI approximation used in NCLE can introduce errors, and that tuning choices are crucial to controlling error and computational cost.

The results reflect tuning behaviour discussed in Section \ref{sec:batch_length}.
Increasing $N$ generally improves NCLE performance.
For most parameters, large $l$ produces worse performance for point estimates and for CI width, reflecting difficulty in estimating high dimensional $\mathbf{x}_b$.
We also found that computation costs increased with $l$.
In this example, the increase is mostly driven by higher costs in training.

Section \ref{sec:batch_length} suggested using a simulation study to tune $l$.
The NCLE results above support using $l=10$,
which gives reasonable coverage and good MCLE accuracy
(although for analytical CL a better choice might be $l=20$ or $100$).
Using Table \ref{tab:garch_times}, we can see that inference with $l=10$ is faster than $l=1000$ by a factor of roughly 4.
Using $l=10$ is also a good match to the heuristic choice from Section \ref{sec:GARCH_batch_length}.
One caveat is that the empirical autocorrelation function suggested using $l=2$, which has worse performance in our simulation study.
So in this case, making a heuristic choice of $l$ using empirical autocorrelations gives roughly the right magnitude for $l$, but a full simulation study is needed to pick the best value.
We make a general recommendation along these lines in Section \ref{sec:discussion}.

\subsubsection{Long sequence experiment} \label{sec:long_seq}

Having concluded that $l=10$ was a good choice of batch length that gave a computational speed-up and good quality inference, we repeat the NCLE experiment with $l=10$ on a longer sequence of length $L=10^6$, equivalent to $B=10^5$ batches. We train NLE using $N=10^5$ and $N=10^6$ training samples.

\begin{figure}[tbp]
    \centering
    \includegraphics[width=0.5\linewidth]{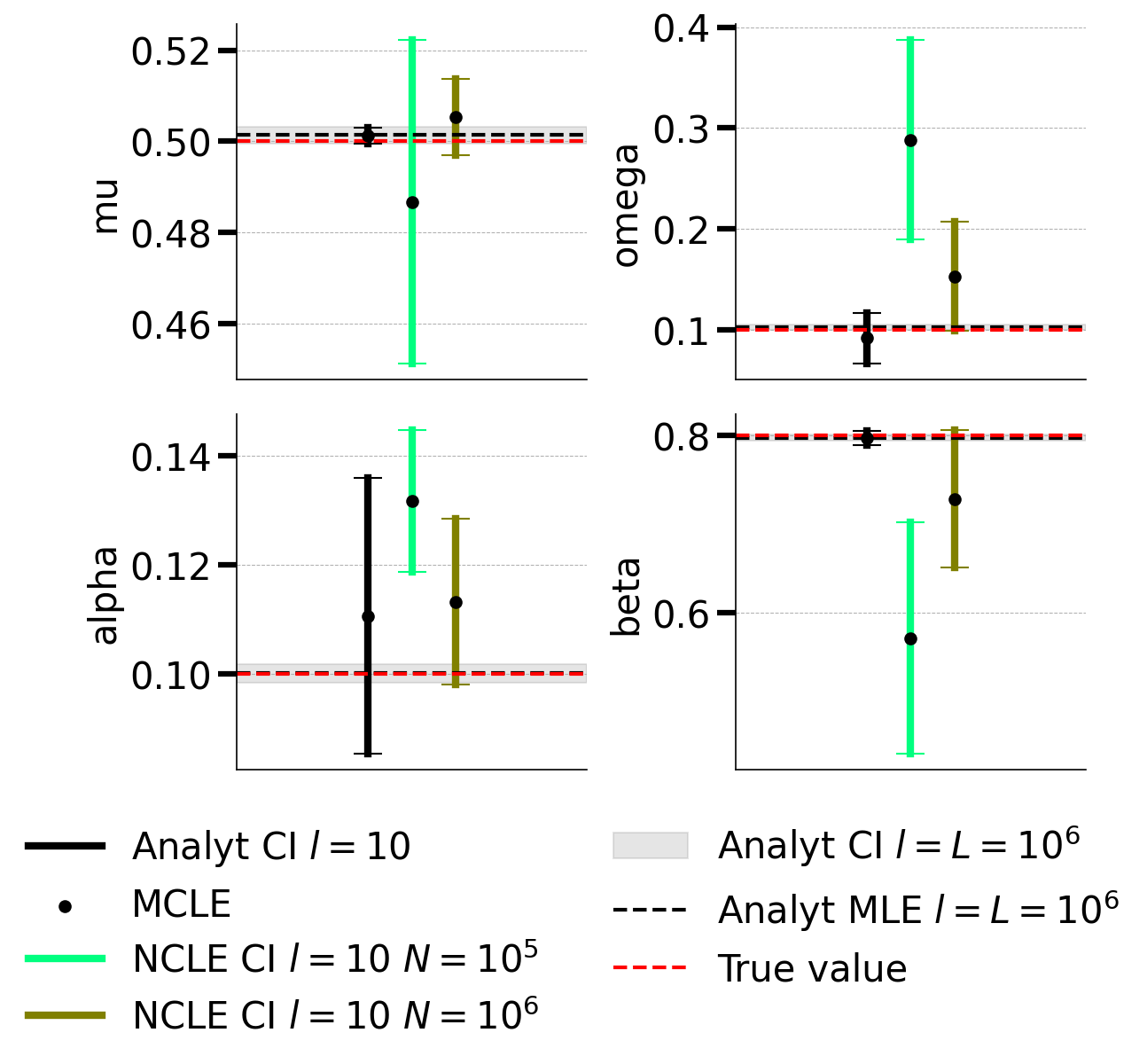}
    \caption{GARCH(1,1) model, $L=10^6$. 95\% confidence intervals.}
    \label{fig:garch_long_ci}
\end{figure}

Figure \ref{fig:garch_long_ci} shows NCLE 95\% confidence intervals obtained for a single observed dataset.
The analytical confidence intervals for $l=L$ were obtained using the Python \texttt{arch} package \citep{arch}.
The analytical CL and $N=10^6$ intervals all contain the true parameter values,
while only 1 of the 4 intervals for $N=10^5$ does so.

The analytical CL intervals for $L=10^6$ are much narrower than for $L=1000$, matching the expected asymptotic behaviour as the sample size increases.
Unlike in the $L=1000$ experiment, the NCLE CIs are notably wider than their analytical CL counterparts, although larger $N$ generally reduces width.
Exploring the reasons for this is an interesting topic for future research.

The NCLE results took a comparable time to the short sequence experiment: a few hours.
However in this case CMLE with \texttt{arch} was much faster, taking only seconds to run.
This illustrates that the NCLE method scales well with sequence length,
but is not competitive with likelihood-based methods where they are available.
Nonetheless, SBI methods can potentially be useful in settings where CMLE is not possible
e.g.~if there are gaps in the data or heavy tailed observations.

Figure \ref{fig:garch_long_ci_noburnin} in Appendix \ref{sec:supp_results} repeats the experiment without
any burn-in in the simulator.
Now all the $N=10^5$ intervals contain the true parameter values, along with 3 of the 4 intervals for $N=10^6$.
The interval lengths are also of a similar scale to when using burn-in.
Since the results have not obviously degraded,
this suggests that using burn-in is not critical for this example.

\section{Discussion} \label{sec:discussion}

We introduce neural composite likelihood estimation (NCLE),
and illustrate its performance on two models.
This shows that, under properly selected batch lengths,
it can produce accurate parameter estimates and confidence intervals, while speeding up inference compared to standard NLE
(by a factor of 16 for the AR example, and 4 for the GARCH example).

We've shown that tuning is crucial to getting good results,
especially the batch length $l$ and number of training simulations $N$.
We recommend using a simulation study to select these,
and outline how this can be carried out.
We also show that making a heuristic choice using empirical autocorrelations can give roughly the right magnitude for $l$,
but a full simulation study is recommended to pick the ideal value.

In exploratory work, we found that SBI sometimes trains poorly.
So in practice, it is likely to be beneficial to make multiple training replications and select the best.

Our likelihood estimate can be based on NLE or NPE.
As explained in Section \ref{sec:batch_length},
we expected that, on raw data, NLE would be better for small $l$ and NPE would allow larger $l$,
but could still degrade for very large $l$.
We investigated this in our AR(1) example, and found that NLE performed well for smaller $l$ values,
while NPE performed well for a middle range of $l$ values.
Interestingly, for $l>100$ both NPE and NLE had problems,
although NPE did still produce reasonable coverage.
So overall for this example, NLE has good performance under a wider range of $l$ than NPE.

An alternative is using NLE with summary statistics,
which can allow larger $l$ and potentially be cheaper due to reduced training time.
We found using sufficient statistics for the AR(1) model produced excellent results.
However in most applications, sufficient statistics are not available,
so selecting informative summary statistics is a crucial tuning choice.

For the GARCH example, our method improves on standard NLE.
However it does not beat using a likelihood-based method (CMLE),
which also gives accurate results, but more quickly.
Nonetheless, this illustrates that NCLE can potentially be useful where likelihood-based methods are not available.
For GARCH, this could include variations with gaps in the data, or heavy tailed observations.

The remainder of this section describes limitations and opportunities for future work.

\subsection{Limitations}

The stationarity assumptions described in Section \ref{sec:NCLE} seem fundamental to our divide-and-conquer approach:
they permit simulation of short batches of data which come from the same distribution as the observed data.
Some mild relaxations seem possible. For instance if the observed sequence does not start in the stationary regime, this should not affect asymptotic behaviour
(as asymptotically almost all data is from the stationary regime.)
However, non-stationary settings are likely to need fundamentally different methods.

Another issue is limited theoretical support for the composite likelihood asymptotics.
More results here would be helpful.
In particular, understanding the effect of the batch length $l$
(similar to the results of \citealp{andrieu2005online} for the SSM setting)
could provide practical guidance on its choice,
and reduce the need for an expensive simulation study.

Our confidence intervals are based on estimating the Godambe matrix
using the results described in Section \ref{sec:ci}.
Their derivation involves assuming that Bartlett identities hold,
which may not be true when using SBI approximations of the likelihood.

\subsection{Future work}

Throughout we've applied SBI using the \texttt{sbi} package defaults for simplicity.
Results could be improved by adjusting these, or using SBI based on other generative approaches such as diffusion models \citep{sharrock2024sequential}.
Also, beyond our investigation of NLE and NPE, other SBI methods could be used.
For instance, neural ratio estimation (NRE) \citep{hermans2019likelihood} estimates the likelihood
up to a constant proportionality depending on the data.
Similarly, variations of composite likelihood could be used.
We could consider overlapping batches or conditional events.
Alternatively, the terms in \eqref{eq:split} could be weighted.
This can balance contributions from unequal batches
e.g.~for datasets with missing values.

We demonstrated that NCLE can be performed using summary statistics, following \eqref{eq:ncl_eq2}.
Future work could investigate methods of learning good summaries specifically for use in NCLE.

The Godambe matrix could be estimated in different ways.
For example, an estimate of $\mathbf{S}$ could be based on evaluating the Hessian in \eqref{eq:S_def} using automatic differentiation.
Also, an estimate of $\mathbf{V}$ can be obtained by using bootstrapped resampling of the observed data in \eqref{eq:V_def}
\citep{coffman2016computationally}.
These and other approaches could reduce reliance on assuming that Bartlett identities hold.

There are also alternative ways to estimate confidence intervals, which avoid estimating the Godambe matrix.
In exploratory work we found these were difficult to tune, but they could be interesting for future work.
One approach is bootstrap confidence intervals \citep{beeravolu2018able}.
We tried the block bootstrap \citep{carlstein1986use},
but found it difficult to tune the block length alongside the batch length $l$ of our method.
Composite likelihood correction methods have been proposed which allow approximate Bayesian inference \citep{pauli2011bayesian, ribatet2012bayesian}.
However again we found it difficult to tune these.

We've concentrated on composite likelihood methods for sequence data models.
Similar approaches could be attempted for other applications e.g.~spatial statistics \citep{maceda2026demonstrating} and point processes \citep{stockman2024sb}.
It would also be interesting to provide a sequential version of NCLE, similar to sequential NLE \citep{papamakarios2019snle}.
The idea is to iteratively use the current likelihood approximation to generate further training data and then improve the approximation.

\bibliography{refs}

\appendix

\section{Composite likelihood regularity conditions} \label{app:regularity}

This appendix expands on Section \ref{sec:asymptotics}
with more details of theory supporting the asymptotic normality MCLE results \eqref{eq:asymptoticMCLE}, and which regularity conditions are required.

\citet{andrieu2005online} prove \eqref{eq:asymptoticMCLE} for state space models.
In particular see equations (41)--(43) in their supplementary material.
Several regularity conditions are required,
including that the model should be stationary and the hidden state should be ergodic.

The AR(1) model of Section \ref{sec:AR1} can be viewed as a state space model
with hidden state $x_t$ and observation $y_t = x_t$.
The GARCH(1,1) model of Section \ref{sec:GARCH} can also be interpreted as a state space model: view $(x_t, \sigma^2_t)$ as a hidden state at time $t$,
which evolves as a Markov chain.
The corresponding observation at time $t$ is $y_t = x_t$.

Both models fail to meet a regularity condition of \citet{andrieu2005online},
requiring a compact parameter space.
Furthermore, the AR(1) model doesn't meet another regularity condition,
requiring $p(x_{t+1} | x_t, \theta)$ to have a non-zero lower bound.

It seems likely that the asymptotic normality results are robust to these minor violations of the regularity conditions.
Our experiments allow us to check this empirically.

\section{AR(1) derivations} \label{sec:ar1_deriv}

This section contains derivations for the AR(1) model defined in Section \ref{sec:AR1}.
We derive sufficient statistics,
and compare asymptotic curvature of the likelihood and composite likelihood.
We consider $x_t$ from the model with $1 \leq t \leq L$ and parameter $\theta$.
Recall that the model is stationary for $\theta \in (-1,1)$ and in this case $\E[x_t^2]=(1-\theta^2)^{-1}$.

The log-likelihood is:
\begin{align*}
    \ell(\theta; \mathbf{x}) &= -\frac{1}{2} \log(2\pi) + \frac{1}{2}\log(1-\theta^2) - \frac{x_1^2(1-\theta^2)}{2}
    + \sum_{t=2}^L \left[ -\frac{1}{2} \log(2\pi)-\frac{(x_t-\theta x_{t-1})^2}{2} \right] \\
    &= -\frac{L}{2} \log(2\pi) + \frac{1}{2}\log(1-\theta^2)  - \frac{1}{2} \sum_{t=1}^L x_t^2 + \theta\sum_{t=2}^L x_{t-1}x_t - \frac{\theta^2}{2}\sum_{t=2}^{L-1} x_t^2.
\end{align*}
(We assume $L>2$. For $L=2$, the first summary vanishes.)
So the sufficient statistics are
\[
    \mathcal{S}_2 = \left(\sum_{t=2}^{L-1} x_t^2, \quad \sum_{t=2}^L x_{t-1}x_t \right).
\]
This differs from typical results for the AR(1) model which have three sufficient statistics (see e.g.~\citealp{forchini2000density}).
This is because we fix $\Var(\epsilon_t) = 1$, rather than using it as an unknown parameter, and we assume $x_1$ follows the stationary distribution.

Differentiating the log-likelihood gives
\begin{align*}
    \ell''(\theta; \mathbf{x}) &= -\frac{1+\theta^2}{(1-\theta^2)^2} - \sum_{t=2}^{L-1} x_t^2.
\end{align*}
Hence
\begin{align*}
-\frac{1}{L} \ell''(\theta;\mathbf{x})
&= \frac{1}{L}\frac{1+\theta^2}{(1-\theta^2)^2} + \frac{1}{L}\sum_{t=2}^{L-1} x_t^2 \\
&= (1-\theta^2)^{-1} + o(1).
\end{align*}
The last line follows using the ergodic theorem
and is true almost surely.

Similarly, our composite log-likelihood is:
\begin{align*}
\ell_c(\theta; \mathbf{x}) =& \sum_{b=1}^B \bigg\{ -\frac{1}{2} \log(2\pi) + \frac{1}{2}\log(1-\theta^2) - \frac{x_{b,1}^2(1-\theta^2)}{2} \\
&+ \sum_{t=2}^l \left[ -\frac{1}{2} \log(2\pi)-\frac{1}{2}(x_{b,t}-\theta x_{b,t-1})^2 \right] \bigg\},
\end{align*}
where $x_{b,t}=x_{(b-1)l+t}$ (entry $t$ in batch $b$).
Differentiating gives
\[
\ell_c''(\theta; \mathbf{x}) = \sum_{b=1}^B \left[ -\frac{1+\theta^2}{(1-\theta^2)^2} - \sum_{t=2}^{l-1} x_{b,t}^2 \right].
\]
Hence:
\begin{align*}
-\frac{1}{L} \ell_c''(\theta; \mathbf{x})
&= \frac{B}{L} \left[ \frac{1+\theta^2}{(1-\theta^2)^2} \right] + \frac{1}{L} \sum_{t=2}^{l-1} \sum_{b=1}^B x_{(b-1)l+t}^2  \\
&= \frac{1}{l} \left[ \frac{1+\theta^2}{(1-\theta^2)^2} + \frac{l-2}{1-\theta^2} \right] + o(1) \\
&= \frac{3\theta^2-1}{l(1-\theta^2)^2} + \frac{1}{1-\theta^2} + o(1).
\end{align*}
The second line above uses the ergodic theorem,
and the result is true almost surely.

\begin{figure}[tbp]
    \centering
    \includegraphics[width=0.6\linewidth]{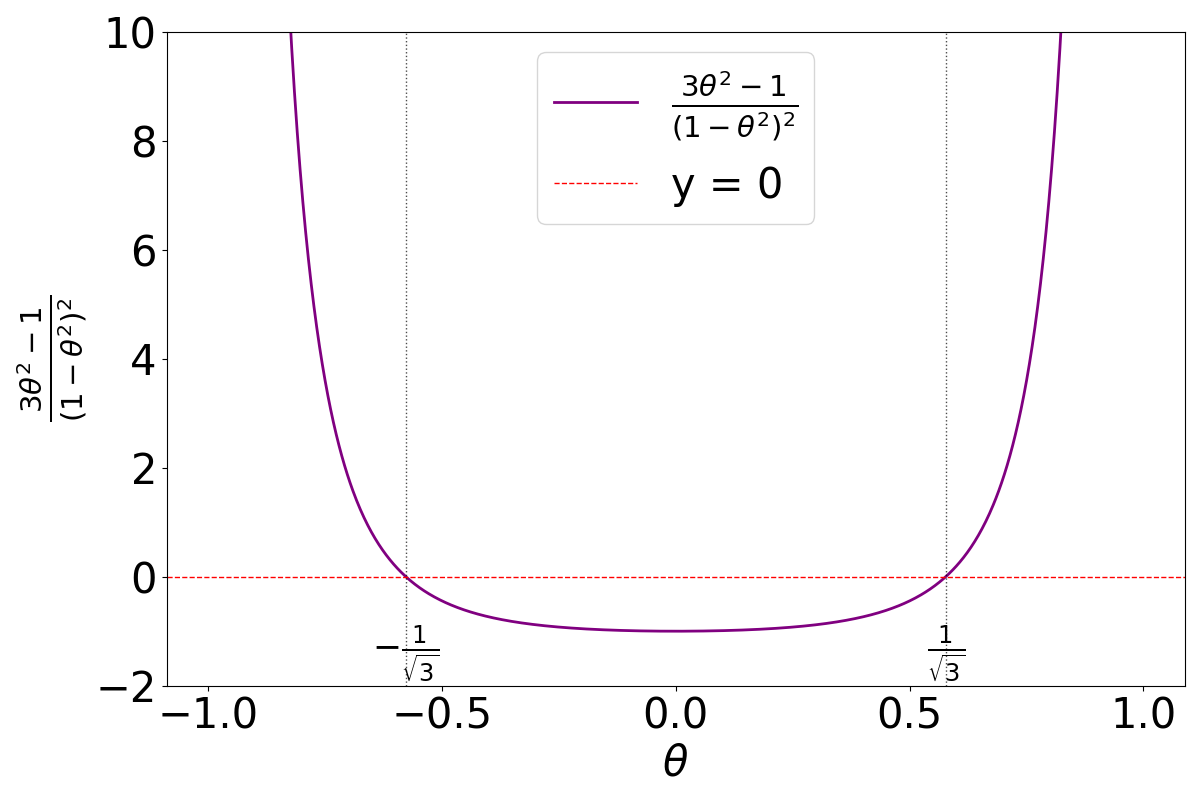}
    \caption{Plot of $\frac{3\theta^2-1}{(1-\theta^2)^2}$. Negative values mean that the composite likelihood has lower asymptotic curvature.}
    \label{fig:ar1_J_vs_Jc}
\end{figure}

From the above,
$\frac{1}{L}[\ell''(\theta) - \ell_c''(\theta)] = \frac{3\theta^2-1}{l(1-\theta^2)^2} + o(1)$
almost surely.
Thus the composite likelihood has lower asymptotic curvature for $\theta \in (-\frac{1}{\sqrt{3}}, \frac{1}{\sqrt{3}})$
(illustrated in Figure \ref{fig:ar1_J_vs_Jc}).
The difference is larger for small $l$.

\section{Supplementary tables and figures}
\label{sec:supp_results}

\begin{figure}[tbp]
    \centering
    \includegraphics[width=0.5\linewidth]{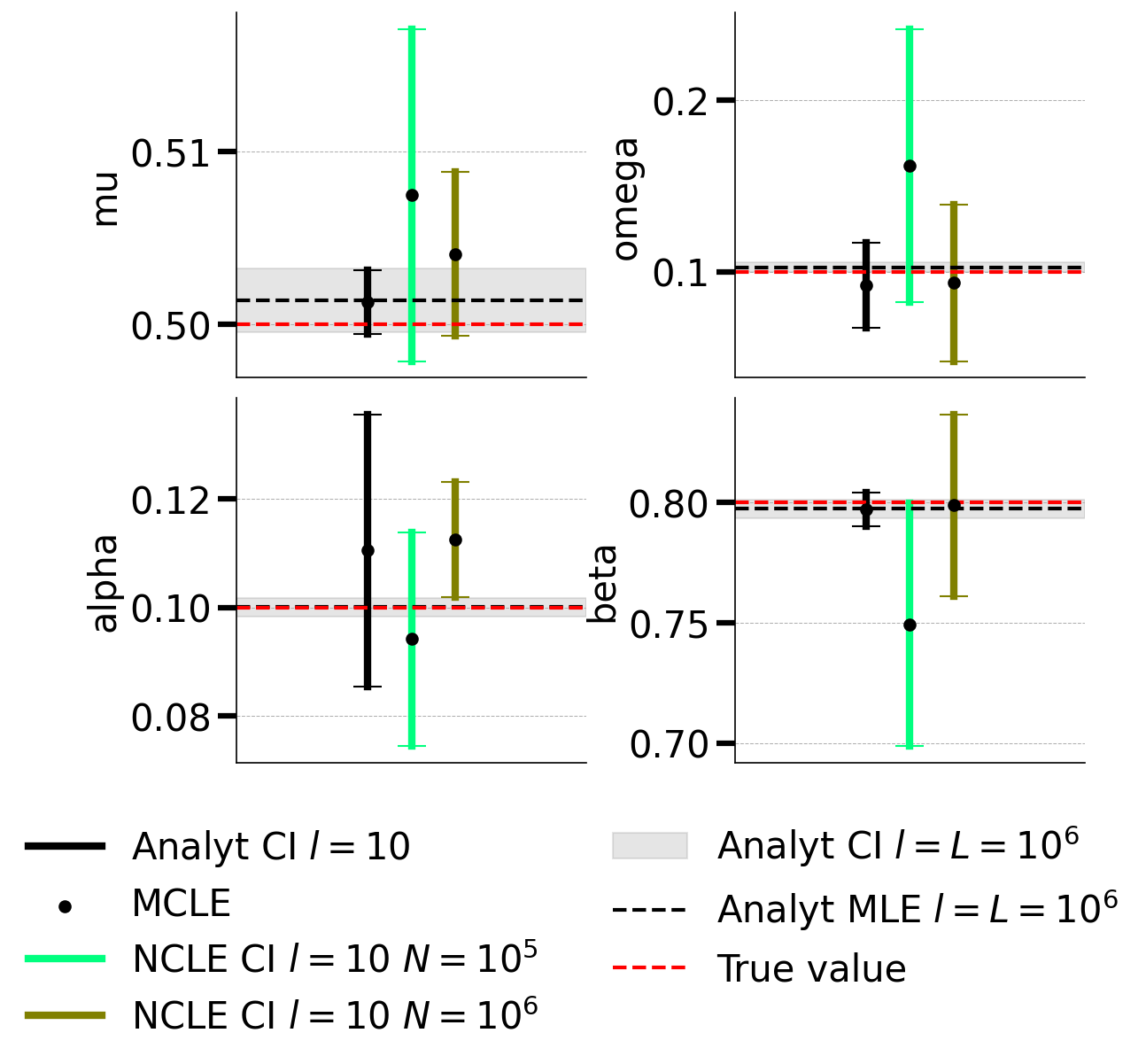}
    \caption{GARCH(1,1) model, $L=10^6$. 95\% confidence intervals. No burn-in.}
    \label{fig:garch_long_ci_noburnin}
\end{figure}

This section mainly contains tables with more numerical details of results shown as plots in the main paper.
There is also an extra plot -- Figure \ref{fig:garch_long_ci_noburnin} --
showing a variation on results from the main paper -- Figure \ref{fig:garch_long_ci} --
but without any burn-in for GARCH simulation.


\begin{table}[htbp]
    \centering
    \begin{tabular}{c|c|c|c|c|c}
    \hline
    \textbf{$l$ ($B$)} & $N=10^4$ & $N=10^5$ & $\mathcal{S}_2$ & NPE & CL \\ \hline
    2 (500)    & 0.106 & 0.087 & 0.878 & 0.279 & 0.080 \\
    5 (200)    & 0.173 & 0.089 & 0.090 & 0.207 & 0.078 \\
    10 (100)   & 0.088 & 0.076 & 0.074 & 0.101 & 0.078 \\
    20 (50)    & 0.096 & 0.072 & 0.139 & 0.087 & 0.077 \\
    100 (10)   & 0.064 & 0.070 & 0.084 & 0.078 & 0.075 \\
    500 (2)    & 0.033 & 0.042 & 0.075 & 0.200 & 0.075 \\
    1000 (1)   & 0.023 & 0.026 & 0.081 & 0.294 & 0.075 \\
    \hline
    \end{tabular}
    \caption{AR(1) model. Mean confidence interval widths. Columns match Figure \ref{fig:ar1_ci_widths} labels.}
    \label{tab:ar1_ci_widths}
\end{table}

\begin{table}[htbp]
    \centering
    \begin{tabular}{c|c|c|c|c}
    \hline
    \textbf{$l$ ($B$)} & Simulations & Training & C.I. calculation & Total \\ \hline
    2 (500)    & 0.004 & 800.70   & 17.09 & 817.79   \\
    5 (200)    & 0.009 & 1282.24  & 14.93 & 1297.18  \\
    10 (100)   & 0.017 & 1118.16  & 14.82 & 1133.00  \\
    20 (50)    & 0.071 & 2885.32  & 14.73 & 2900.12  \\
    100 (10)   & 0.325 & 4998.93  & 15.08 & 5014.34  \\
    500 (2)    & 1.727 & 17794.77 & 14.09 & 17810.59 \\
    1000 (1)   & 1.955 & 17767.86 & 14.23 & 17784.05 \\ \hline
    \end{tabular}
    \caption{AR(1) model. Mean times in seconds for raw data NLE with $N=10^5$. Columns match Figure \ref{fig:ar1_times} labels.}
    \label{tab:ar1_times}
\end{table}

\begin{table}[tbp]
    \centering
    \begin{tabular}{c|c|c|c|c|c}
    \hline
    \textbf{$l$ ($B$)} & $N=10^4$ & $N=10^5$ & $\mathcal{S}_2$ & NPE & CL \\ \hline
    2 (500)    & 87\%   & 93\%   & 87.5\%  & 98\%   & 93.5\% \\
    5 (200)    & 97.5\% & 91\%   & 97\%    & 97.5\% & 94.5\% \\
    10 (100)   & 93\%   & 96\%   & 89.5\%    & 86\%   & 93.5\% \\
    20 (50)    & 82\%   & 92\%   & 95.5\%    & 86.5\% & 91.5\% \\
    100 (10)   & 68\%   & 88.5\% & 93.5\%    & 88\%   & 92.5\% \\
    500 (2)    & 35\%   & 49.5\% & 93.5\%  & 95.5\% & 96\% \\
    1000 (1)   & 20\%   & 41\%   & 92.5\%   & 93.5\% & 95.5\% \\
    \hline
    \end{tabular}
    \caption{AR(1) model. 95\% coverages. Columns match Figure \ref{fig:ar1_coverages} labels.}
    \label{tab:ar1_coverages}
\end{table}


\begin{table}[htbp]
    \centering
    \begin{tabular}{c|c|c|c|c}
    \hline
    \textbf{$l$ ($B$)} & Simulations & Training & C.I. calculation & Total \\ \hline
    2 (500)    & 37.38 & 7777.35  & 15.78 & 7830.50 \\
    5 (200)    & 36.10 & 9126.42  & 15.44 & 9177.96 \\
    10 (100)   & 34.04 & 14047.92 & 14.66 & 14096.62 \\
    20 (50)    & 40.18 & 17460.94 & 14.88 & 17516.00 \\
    100 (10)   & 29.24 & 16122.88 & 16.47 & 16168.58 \\
    500 (2)    & 74.74 & 47286.64 & 26.32 & 47387.69 \\
    1000 (1)   & 108.22 & 54810.43 & 35.82 & 54954.47 \\ \hline
    \end{tabular}
    \caption{GARCH(1,1) model. Mean times in seconds with $N=10^6$. Columns match Figure \ref{fig:garch_times} labels.}
    \label{tab:garch_times}
\end{table}

\begin{table}[tbp]
    \centering
    \renewcommand{\arraystretch}{0.7}
    \begin{tabular}{llccccccc}
    \toprule
    & & \multicolumn{7}{c}{\textbf{$l$ ($B$)}} \\
    \cmidrule(lr){3-9}
    & & 2 (500) & 5 (200) & 10 (100) & 20 (50) & 100 (10) & 500 (2) & 1000 (1) \\
    \midrule
    \multirow{4}{*}{\shortstack[l]{NCLE\\$N=10^4$}}
    & $\mu$    & 0.404 & 0.126 & 0.141 & 0.149 & 0.127 & 0.143 & 0.138 \\
    & $\omega$ & 0.415 & 0.351 & 0.694 & 0.418 & 0.352 & 0.347 & 0.314 \\
    & $\alpha$ & 0.400 & 0.128 & 0.218 & 0.258 & 0.205 & 0.208 & 0.193 \\
    & $\beta$  & 0.406 & 0.286 & 0.347 & 0.424 & 0.306 & 0.313 & 0.285 \\
    \midrule
    \multirow{4}{*}{\shortstack[l]{NCLE\\$N=10^5$}}
    & $\mu$    & 0.186 & 0.128 & 0.141 & 0.119 & 0.118 & 0.113 & 0.128 \\
    & $\omega$ & 0.567 & 0.355 & 0.325 & 0.319 & 0.367 & 0.568 & 0.830 \\
    & $\alpha$ & 0.238 & 0.198 & 0.247 & 0.275 & 0.229 & 0.363 & 0.673 \\
    & $\beta$  & 0.573 & 0.459 & 0.455 & 0.381 & 0.487 & 0.846 & 1.282 \\
    \midrule
    \multirow{4}{*}{\shortstack[l]{NCLE\\$N=10^6$}}
    & $\mu$    & 0.127 & 0.176 & 0.121 & 0.120 & 0.131 & 0.115 & 0.107 \\
    & $\omega$ & 0.957 & 0.470 & 0.359 & 0.327 & 0.320 & 0.508 & 0.697 \\
    & $\alpha$ & 0.359 & 0.159 & 0.168 & 0.188 & 0.152 & 0.259 & 0.502 \\
    & $\beta$  & 0.964 & 0.538 & 0.468 & 0.464 & 0.406 & 0.690 & 1.135 \\
    \midrule
    \multirow{4}{*}{CL}
    & $\mu$    & -     & - & 0.139 & 0.117 & 0.116 & 0.119 & 0.118 \\
    & $\omega$ & -     & -     & 0.501 & 0.264 & 0.215 & 0.210 & 0.209 \\
    & $\alpha$ & -     & - & 0.168 & 0.127 & 0.115 & 0.115 & 0.119 \\
    & $\beta$  & -     & -     & 0.597 & 0.344 & 0.292 & 0.284 & 0.287 \\
    \bottomrule
    \end{tabular}
    \caption{GARCH(1,1) model. Confidence interval widths. A dash denotes that the value is absent due to numerical problems in calculation.}
    \label{tab:garchCIwidths}
\end{table}

\begin{table}[tbp]
\centering
\renewcommand{\arraystretch}{0.9}
\makebox[\textwidth][c]{ 
\begin{tabular}{lccccccc}
\toprule
& \multicolumn{7}{c}{\textbf{$l$ ($B$)}} \\
\cmidrule(lr){2-8}
& 2 (500) & 5 (200) & 10 (100) & 20 (50) & 100 (10) & 500 (2) & 1000 (1) \\
\midrule
$N=10^4$ & -22.08 & -23.61 & -21.93 & -21.37 & -23.52 & -23.48 & -24.87 \\
$N=10^5$ & -21.02 & -22.94 & -22.52 & -22.76 & -23.55 & -23.72 & -21.75 \\
$N=10^6$ & -19.74 & -22.61 & -23.86 & -23.81 & -23.96 & -25.00 & -24.23 \\
CL        & -      & -      & -22.70 & -24.74 & -25.22 & -25.21 & -25.08 \\
\bottomrule
\end{tabular}
}
\caption{Mean $\log\det(\mathbf{G}^{-1})$ values for GARCH(1,1) model, corresponding to Figure \ref{fig:garch_det}. More negative values indicate greater precision. A dash denotes an absent value due to numerical problems in calculation.
(We include $N=10^4$ results not shown in Figure \ref{fig:garch_det} to avoid cluttering the plot.)}
\label{tab:garch_det}
\end{table}

\begin{table}[tbp]
    \centering
    \renewcommand{\arraystretch}{0.7}
    \begin{tabular}{llccccccc}
    \toprule
    & & \multicolumn{7}{c}{\textbf{$l$ ($B$)}} \\
    \cmidrule(lr){3-9}
    & & 2 (500) & 5 (200) & 10 (100) & 20 (50) & 100 (10) & 500 (2) & 1000 (1) \\
    \midrule
    \multirow{4}{*}{\shortstack[l]{NCL\\$N=10^4$}}
    & $\mu$    & 96\%   & 94\%   & 90.5\% & 93\%   & 87.5\% & 88.5\% & 66\% \\
    & $\omega$ & 26.5\% & 8.5\%  & 28\%   & 27\%   & 7.5\%  & 6\%    & 11\% \\
    & $\alpha$ & 78.5\% & 60.5\% & 74.5\% & 77.5\% & 66\%   & 62\%   & 60\% \\
    & $\beta$  & 70.5\% & 9\%    & 24\%   & 38\%   & 14\%   & 11\%   & 21\% \\
    \midrule
    \multirow{4}{*}{\shortstack[l]{NCL\\$N=10^5$}}
    & $\mu$    & 93.5\% & 93\%   & 93\%   & 88.5\% & 93.5\% & 90.5\% & 90.5\% \\
    & $\omega$ & 46\%   & 57\%   & 45.5\% & 78.5\% & 81.5\% & 79\%   & 70\% \\
    & $\alpha$ & 95.5\% & 78.5\% & 81.5\% & 94\%   & 90.5\% & 77\%   & 67.5\% \\
    & $\beta$  & 44.5\% & 55.5\% & 53\%   & 78\%   & 81.5\% & 78\%   & 63\% \\
    \midrule
    \multirow{4}{*}{\shortstack[l]{NCL\\$N=10^6$}}
    & $\mu$    & 94\%   & 95.5\% & 94\%   & 92.5\% & 90.5\% & 93.5\% & 85\% \\
    & $\omega$ & 89.5\% & 75.5\% & 77.5\% & 86.5\% & 83\%   & 75.5\% & 82\% \\
    & $\alpha$ & 97.5\% & 95\%   & 95\%   & 91.5\% & 88.5\% & 83.5\% & 86\% \\
    & $\beta$  & 85\%   & 75.5\% & 78\%   & 82.5\% & 81.5\% & 81\%   & 85\% \\
    \midrule
    \multirow{4}{*}{CL}
    & $\mu$    & - & - & 93.5\% & 94.5\% & 94\%   & 93\%   & 93.5\% \\
    & $\omega$ & - & -   & 97.5\% & 90.5\% & 91.5\% & 93.5\% & 96\% \\
    & $\alpha$ & - & -   & 91.5\% & 90.5\% & 95\%   & 93\%   & 95.5\% \\
    & $\beta$  & - & -   & 94.5\% & 93\%   & 92.5\% & 93.5\% & 95.5\% \\
    \bottomrule
    \end{tabular}
\caption{GARCH(1,1) model 95\% coverages. A dash denotes an absent value due to numerical problems in calculation.}
\label{tab:garch_coverages_combined}
\end{table}

\end{document}